\documentclass[twocolumn]{aastex62}

\usepackage{amsmath}
\usepackage{color}

\newcommand\HST{{\it HST }}

\received{2019}
\revised{2020}
\accepted{2020}

\shorttitle{Late-time afterglows of GRB 160625B and GRB 160509A}
\shortauthors{Kangas et al.}

\begin{document}

\title{The late-time afterglow evolution of long gamma-ray bursts GRB 160625B and GRB 160509A}


\correspondingauthor{Tuomas Kangas}
\email{tkangas@stsci.edu}

\author[0000-0002-5477-0217]{Tuomas Kangas}
\affil{Space Telescope Science Institute, 3700 San Martin Drive, Baltimore, MD 21218, USA}

\author{Andrew S. Fruchter}
\affil{Space Telescope Science Institute, 3700 San Martin Drive, Baltimore, MD 21218, USA}

\author{S. Bradley Cenko}
\affil{Astrophysics Science Division, NASA Goddard Space Flight Center, Mail Code 661, Greenbelt, MD 20771, USA}
\affil{Joint Space-Science Institute, University of Maryland, College Park, MD 20742, USA}

\author{Alessandra Corsi}
\affil{Department of Physics and Astronomy, Texas Tech University, Box 1051, Lubbock, TX 79409-1051, USA}

\author{Antonio de Ugarte Postigo}
\affil{Instituto de Astrof\'{i}sica de Andaluc\'{i}a, Glorieta de la Astronom\'{i}a s/n, E-18008, Granada}
\affil{Dark Cosmology Centre, Niels Bohr Institute, Juliane Maries Vej 30, Copenhagen {\O}, 2100, Denmark}

\author{Asaf Pe'er}
\affil{Department of Physics, Bar-Ilan University, Ramat-Gan 52900, Israel}

\author{Stuart N. Vogel}
\affil{Department of Astronomy, University of Maryland, College Park, MD 20742, USA}

\author{Antonino Cucchiara}
\affil{University of Virgin Islands, College of Science and Mathematics, \#2 Brewers Bay Road, Charlotte Amalie, USV1 00802}

\author{Benjamin Gompertz}
\affil{Department of Physics, University of Warwick, Coventry, CV4 7AL, UK}

\author{John Graham}
\affil{Kavli Institute for Astronomy and Astrophysics, Peking University, Beijing 100871, PR China}

\author{Andrew Levan}
\affil{Department of Physics, University of Warwick, Coventry, CV4 7AL, UK}

\author{Kuntal Misra}
\affil{Aryabhatta Research Institute of Observational Sciences (ARIES), Manora Peak, Nainital 263 002, India}

\author{Daniel A. Perley}
\affil{Astrophysics Research Institute, Liverpool John Moores University, IC2, Liverpool Science Park, \\ 146 Brownlow Hill, Liverpool L3 5RF, UK}

\author{Judith Racusin}
\affil{Astrophysics Science Division, NASA Goddard Space Flight Center, Mail Code 661, Greenbelt, MD 20771, USA}

\author{Nial Tanvir}
\affil{University of Leicester, Department of Physics \& Astronomy and Leicester Institute of Space \& Earth Observation \\
University Road, Leicester, LE1 7RH UK}



\begin{abstract}

We present post-jet-break \textit{HST}, VLA and \textit{Chandra} observations of the afterglow of the long $\gamma$-ray bursts GRB 160625B (between 69 and 209 days) and GRB 160509A (between 35 and 80 days). We calculate the post-jet-break decline rates of the light curves, and find the afterglow of GRB 160625B inconsistent with a simple $t^{-3/4}$ steepening over the break, expected from the geometric effect of the jet edge entering our line of sight. However, the favored optical post-break decline ($f_{\nu} \propto t^{-1.96 \pm 0.07}$) is also inconsistent with the $f_{\nu} \propto t^{-p}$ decline (where $p \approx 2.3$ from the pre-break light curve), which is expected from exponential lateral expansion of the jet; perhaps suggesting lateral expansion that only affects a fraction of the jet. The post-break decline of GRB 160509A is consistent with both the $t^{-3/4}$ steepening and with $f_{\nu} \propto t^{-p}$. We also use {\sc boxfit} to fit afterglow models to both light curves and find both to be energetically consistent with a millisecond magnetar central engine, although the magnetar parameters need to be extreme (i.e. $E \sim 3 \times 10^{52}$ erg). Finally, the late-time radio light curves of both afterglows are not reproduced well by {\sc boxfit} and are inconsistent with predictions from the standard jet model; instead both are well represented by a single power law decline (roughly $f_{\nu} \propto t^{-1}$) with no breaks. This requires a highly chromatic jet break ($t_{j,\mathrm{radio}} > 10 \times t_{j,\mathrm{optical}}$) and possibly a two-component jet for both bursts.

\end{abstract}



\section{Introduction} \label{sec:intro}

Gamma-ray bursts (GRBs) are among the most luminous transient events in the universe. Through their association with broad-lined type Ic supernovae \citep[e.g.][]{iwamoto98,wbloom06,hjbloom12}, long GRBs (LGRBs; duration of the prompt $\gamma$-ray emission more than 2 s) have been established as the terminal core-collapse explosions of massive stars at cosmological distances \citep[e.g.][]{paczynski86,woosley93,mfwoosley99}, where an ultra-relativistic jet is launched and breaks out of the stellar envelope, generating the initial prompt emission of $\gamma$ rays through an as yet unclear mechanism \citep[for a review on GRB physics, see e.g.][]{piran04, kumarzhang15}. The central engine responsible for launching the jet and powering the emission may be either accretion onto a black hole formed in the core collapse \citep{woosley93} or rotational energy released through the spin-down of a nascent magnetar \citep[e.g.][]{bucciantini08,bucciantini09}. The prompt emission of a GRB is followed by an afterglow from X-ray to radio frequencies -- synchrotron emission from an external shock created by the interaction between the circumburst medium (CBM) and the highly collimated and relativistically beamed jet \citep[e.g.][]{pacz93,sari98,piran04}. The flux density of the afterglow declines as a power law of the form $f_{\nu} \propto t^{\alpha}$. 

As the jet interacts with the CBM, it decelerates and the relativistic beaming effect diminishes over time \citep[on the order of days or weeks after a long GRB; e.g.][]{racusin09}. This results in an achromatic \emph{jet break} in the afterglow light curve when the relativistic beaming angle ($\Gamma^{-1}$, where $\Gamma$ is the bulk Lorentz factor in the jet) becomes comparable to the opening angle of the jet \citep{rhoads99,sari99}, with a steeper power-law decline after the break. The post-break decline is affected by a geometric 'edge effect', in contrast to the
situation pre-break where the observer only sees a fraction of the jet front and hence behaviour consistent with an isotropic fireball model. This phenomenon is believed to steepen the decline slope $\alpha$ by $-3/4$ over the break assuming a constant-density CBM, or by $-1/2$ in the case of a wind-like CBM \citep[e.g.][]{mrees99,panmesz99,kumarzhang15}. Another effect is that, around the same time as this happens, transverse sound waves become able to cross the jet and lateral expansion starts, exponentially decelerating the shock wave. Theoretically the post-break slope in this scenario is expected to be equal to $-p$ \citep[e.g.][]{sari99}, where $p$ is the index of the electron Lorentz factor distribution ($N(\gamma) \propto \gamma^{-p}$), typically estimated to be between 2 and 3. There is, however, evidence from numerical simulations that the lateral expansion is unimportant until a later stage -- at least unless the jet is very narrow, $\theta_j \lesssim 3$ deg \citep{lyutikov2012,granot2012}. At even later times, the jet is expected to be better described as a non-relativistic fireball in the Sedov-von Neumann-Taylor regime, resulting in a somewhat flatter decline \citep[e.g.][]{fwk00,vdh08}.

Simulations of relativistic shocks have resulted in values around $p \approx 2.2$ \citep[e.g.][]{bednarz98,gallant99,kirk2000}. In the X-rays, the pre-break light curve tends to follow a decline around $t^{-1.2}$ \citep[albeit with some variation; e.g.][]{piran04,zhang06}; thus both of these effects result in a roughly similar post-break decline (i.e. $\sim t^{-2}$, though with high uncertainties due to the the fast decline and the resulting faintness; often there are not enough data to distinguish between $t^{-1.9}$ and $t^{-2.2}$). Thus determining the exact scenario observationally requires late-time observations of the rapidly declining afterglows to constrain this slope.

The Large Area Telescope (LAT) on the \textit{Fermi} Gamma-ray Space Telescope has detected a number of GRBs at relatively high energies (MeV to GeV) since the launch of \textit{Fermi} in 2008. These are often among the most energetic GRBs,
consistent with the Amati correlation between isotropic-equivalent energy $E_{\mathrm{iso}}$ and the peak of the energy spectrum \citep{amati02}, and can haveisotropic-equivalent energies on the order of $10^{54}$ erg \citep{cenko11}. Some of these most energetic bursts do not exhibit the expected jet breaks, suggesting larger opening angles than expected and making them even more energetic intrinsically \citep{depasq16,gompertz17}. With beaming-corrected energies on the order of $10^{52}$ erg, magnetar spin-down models struggle to produce the required power \citep{cenko11}. Thus examining the late-time evolution of the LAT bursts can shed light on the physics of the most energetic GRBs.

In this paper, we present results from our late-time \textit{Hubble Space Telescope} (\textit{HST}), Karl G. Jansky Very Large Array (VLA) and \textit{Chandra X-ray Observatory} imaging observations of the afterglows of two LAT bursts, GRB 160625B and GRB 160509A. GRB 160625B was discovered by the Gamma-ray Burst Monitor (GBM) on \textit{Fermi} on 2016 June 25 at 22:40:16.28 UT \citep[MJD 57564.9;][]{160625b_disc} and detected by the LAT as well. \citet{160625b_z} determined its redshift to be $z=1.406$. It was one of the most energetic $\gamma$-ray bursts ever observed with $E_{iso} \sim 3 \times 10^{54}$ erg \citep{wang17,zhang18}, and a well-studied object with a multi-frequency follow-up that revealed signs of a reverse shock within the jet \citep{alexander17}. The jet break time was unusually long, around 20 days, as expected from unusually bright GRBs \citep[the median time is $\sim 1 $ d, with more energetic bursts having longer break times; see][]{racusin09}. GRB 160509A was detected by GBM and LAT on 2016 May 9 at 08:59:04.36 UT \citep[MJD 57517.4;][]{roberts16,longo16,longo16b} at a redshift of $z = 1.17$ \citep{tanvir16}. With $E_{iso} = 8.6\pm1.1\times10^{53}$ erg, this was another luminous burst that exhibited signs of a reverse shock as well \citep{laskar16}. 

Our observations of GRB 160625B make its follow-up one of the longest post-jet-break optical and X-ray follow-ups of a GRB afterglow\footnote{The post-break light curve of GRB 060729 \citep{grupe10} and GRB 170817A \citep{hajela19} has been followed up longer, while GRB 130427A was followed for $\sim1000$ days \citep[][]{depasq16}, but exhibited no jet break.}, thus providing one of the best estimates of the post-break decline in these bands so far, while for GRB 160509A no prior estimates of the infrared/optical post-break decline could be made due to the very sparse light curve. 

Our observations and data reduction process are described in Section \ref{sec:data}. Our analysis and results are presented in Section \ref{sec:analysis}. In Section \ref{sec:disco}, we discuss the implications of our findings, and finally present our conclusions in Section \ref{sec:concl}. All magnitudes are in the AB magnitude system \citep{abmags} and all error bars correspond to $1 \sigma$ confidence intervals. We use the cosmological parameters $H_0 = 69.6$ km s$^{-1}$ Mpc$^{-1}$, $\Omega_m = 0.286$ and $\Omega_\Lambda = 0.714$ \citep{bennett14}.

\section{Observations and data reduction} \label{sec:data}

Late-time imaging observations of GRB 160625B were performed using \textit{HST}/WFC3 and the F606W filter on 2016 September 5 (71.5 d) and 2016 November 13 (140.2 d). A template image of the host galaxy was created by combining images obtained with the same setup on 2017 November 6 (498.3 d) and 11 (503.6 d).  At this time the contribution of the afterglow itself was a factor of $\sim 13$ fainter than at 140 d, assuming a $f_{\nu} \propto t^{\alpha}$ decline where $\alpha = -2$. Imaging of GRB 160509A in the $H$ band was performed using the Canarias InfraRed Camera Experiment \citep[CIRCE;][]{circe} instrument on Gran Telescopio Canarias (GTC) on 2016 May 15 (5.8 d) and 2016 June 3 (24.8 d). Late-time imaging of GRB 160509A was done using \textit{HST}/WFC3 and the F110W and F160W filters on 2016 June 13 (35.3 d); template images of the host galaxy in these filters were obtained on 2017 July 5 (422.1 d), when, assuming $\alpha = -2$, the afterglow was a factor of 143 fainter. Our \textit{HST} observations of both bursts were executed as part of program GO 14353 (PI Fruchter), and these data are available at\dataset[10.17909/t9-yvpg-xb33]{\doi{10.17909/t9-yvpg-xb33}} (GRB 160625B) and\dataset[10.17909/t9-11cx-cv41]{\doi{10.17909/t9-11cx-cv41}} (GRB 160509A).

Basic reduction and flux calibration of the \textit{HST} images was performed by the \textit{HST} {\sc calwf3} pipeline. The calibrated images were corrected for distortion, drizzled \citep{drizzle} and aligned to a common world coordinate system using the \texttt{astrodrizzle}, \texttt{tweakreg} and \texttt{tweakback} tasks in the {\sc drizzlepac}\footnote{\href{http://drizzlepac.stsci.edu/}{http://drizzlepac.stsci.edu/}} package in {\sc pyraf}\footnote{\href{http://www.stsci.edu/institute/software_hardware/pyraf}{http://www.stsci.edu/institute/software\_hardware/pyraf}}. The two epochs of GRB 160625B in November 2017 were combined into one template image. Subtraction of the template images and aperture photometry of the afterglows were done using {\sc iraf}\footnote{{\sc iraf} is distributed by the National Optical Astronomy Observatory, which is operated by the Association of Universities for Research in Astronomy (AURA) under cooperative agreement with the National Science Foundation.}. Basic reduction of the GTC/CIRCE data was done using standard {\sc iraf} tasks. The \textit{HST} F160W template image was subtracted from the CIRCE images using the {\sc isis} 2.2 package \citep{alardlupton, alard}. Flux calibration was done using field stars in the Two-Micron All Sky Survey (2MASS) catalog\footnote{http://www.ipac.caltech.edu/2mass/} \citep{2mass}, and aperture photometry was performed using standard {\sc iraf} tasks. At 24.8 d, we were unable to detect the afterglow and only obtained a ($3\sigma$) limit of $H \geq 21.9$ mag.

\begin{figure}
\includegraphics[width=1.0\columnwidth]{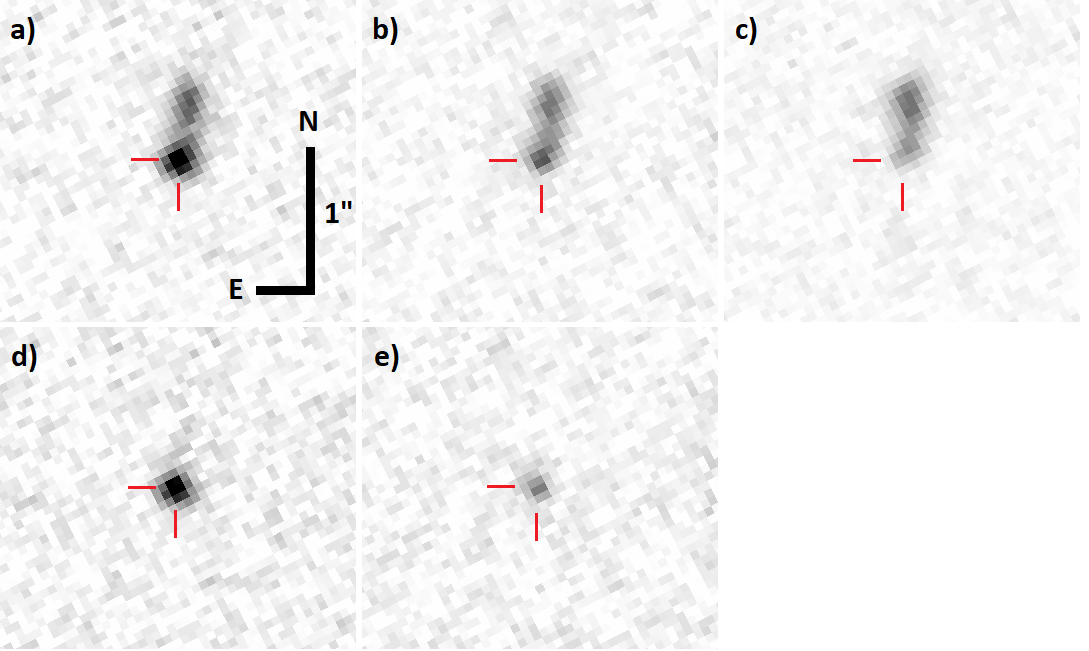} 
\caption{Afterglow and host galaxy of GRB 160625B in the F606W band. \textit{Panel a}: the afterglow and the host galaxy at 71.5 d; \textit{b}: 140.2 d; \textit{c}: the combined template at $\sim500$ d; \textit{d}: the template-subtracted image at 71.5 d; and \textit{e}: the subtraction at 140.2 d. North is up and East is to the left in all panels. The black North-South line corresponds to one arcsecond. The afterglow location is indicated with red tick-marks.}
\label{fig:field}
\end{figure}

\begin{figure}
\includegraphics[width=1.0\columnwidth]{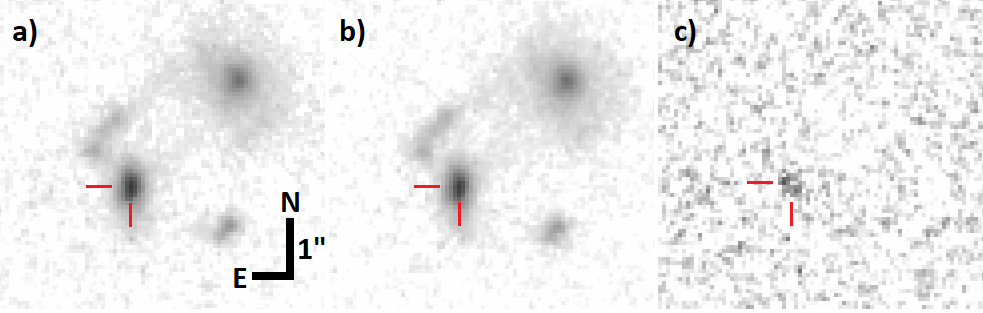} 
\caption{Afterglow and host galaxy of GRB 160509A in the F160W band. \textit{Panel a}: the afterglow and the host galaxy at 35.3 d; \textit{b}: the template at 422.1 d; \textit{c}: the template-subtracted image at 35.3 d. North is up and East is to the left in all panels. The black North-South line corresponds to one arcsecond. The afterglow location is indicated with red tick-marks. The afterglow is very weak compared to the host galaxy, making a template subtraction crucial for this target.}
\label{fig:field2}
\end{figure}

The measured magnitudes of GRB 160625B were corrected for over-subtraction caused by the continued presence of a faint afterglow in the template image. Assuming a post-jet-break decline of $\alpha = -2.0 \pm 0.2$ (obtained from a single-power-law fit to \textit{uncorrected} $>25$~d data, with errors rounded up to be conservative), the afterglow flux present in the template image was estimated to be $2.0\pm1.0$ per cent of the flux at 71.5 d or $7.5\pm2.6$ per cent of the flux at 140.2 d, and thus the images at these epochs were over-subtracted by approximately these amounts. The magnitudes were adjusted for this; the errors of the corrected magnitudes include an estimate of the uncertainty of the over-subtraction. The magnitudes of GRB 160509A were not corrected, as the contribution of the afterglow in the template image was only estimated to be 0.7 per cent of the 35.3 d brightness. The log of optical observations and measured and corrected magnitudes of GRB 160625B are presented in Table \ref{table:phot}, while Table \ref{table:phot_nir} contains the near-infrared observations of GRB 160509A. Figure \ref{fig:field} shows our F606W band images and the resulting template subtractions of GRB 160625B, while Figure \ref{fig:field2} shows the F160W image and subtraction of GRB 160509A.

Late-time X-ray imaging of both GRBs was performed using \textit{Chandra}/ACIS-S in VFAINT mode (proposal ID 17500753, PI Fruchter). GRB 160625B was observed on 2016 September 3 (69.8 d), 2016 November 15 (142.3 d) and 2016 November 19 (146.2 d). The latter two epochs were combined to obtain the flux at $144.3 \pm 2.2$ d, as the flux of the afterglow was not expected to vary significantly over a few days at this time. GRB 160509A was observed on 2016 June 20 (42.1 d). Reprocessing of the Chandra level 1 data was performed using the \texttt{chandra\_repro} script within the {\sc ciao} v. 4.9 software \citep[{\sc caldb} v. 4.7.7;][]{ciao}, and aperture photometry was done using {\sc iraf}. The web-based Portable Interactive Multi-Mission Simulator ({\sc pimms}\footnote{\href{https://heasarc.gsfc.nasa.gov/docs/software/tools/pimms.html}{https://heasarc.gsfc.nasa.gov/docs/software/tools/pimms.html}}) was used to convert count rates in the 0.3 -- 10 keV range to unabsorbed flux densities at 5 keV. For GRB 160625B, we used a Galactic neutral hydrogen column density $N_{\mathrm{H,MW}} = 9.76 \times 10^{20}$ cm$^{-2}$ \citep{column}, a photon index of $\Gamma_X = 1.86$ and an intrinsic absorption of $N_{\mathrm{H,int}} = 2.1 \times 10^{21}$ cm$^{-2}$ as derived by \citet{alexander17}. These parameters are also consistent with the initial analysis by \citet{melandri16}. For GRB 160509A, we used a Galactic neutral hydrogen column density $N_{\mathrm{H,MW}} = 2.12 \times 10^{21}$ cm$^{-2}$ \citep{column}, a photon index of $\Gamma_X = 2.07$ and an intrinsic absorption of $N_{\mathrm{H,int}} = 1.52 \times 10^{22}$ cm$^{-2}$, following \citet{laskar16}. $\Gamma_X$ is assumed to be constant over the light curve break. The log of X-ray observations and derived flux densities is presented in Table \ref{table:xray}.

\begin{table}
\begin{small}
\caption{Log of our late-time \textit{HST}/WFC3 observations of GRB 160625B.}
\begin{tabular}{llccc}
\hline
Phase & MJD & $t_{\mathrm{exp}}$ & F606W & corrected F606W \\
(d) & & (s)  & (mag) & (mag) \\
\hline
71.5 & 57636.4 & 2400 & $25.38 \pm 0.03$ & $25.36 \pm 0.04$\\
140.2 & 57705.1 & 4800 & $26.76 \pm 0.06$ & $26.67 \pm 0.07$\\
498.3 & 58063.2 & 4800 & ... & ... \\
503.6 & 58068.5 & 4800 & ... & ... \\
 \hline
\end{tabular}
\label{table:phot}
\end{small}
\end{table}

\begin{table*}
\centering
\begin{minipage}{0.75\linewidth}
\begin{small}
\caption{Log of our late-time \textit{HST}/WFC3 and GTC/CIRCE observations of GRB 160509A.}
\begin{tabular}{llcccccc}
\hline
Phase & MJD & $t_{\mathrm{exp,F110W}}$ & F110W & $t_{\mathrm{exp,F160W}}$ & F160W & $t_{\mathrm{exp,}H}$ & $H$ \\
(d) & & (s)  & (mag) & (s)  & (mag) & (s) & (mag) \\
\hline
5.8 & 57523.2 & ... & ... & ... & ... & 3060 & $20.50 \pm 0.17$ \\
24.8 & 57542.2 & ... & ... & ... & ... & 2100 & $\geq 21.9$ \\
35.3 & 57552.7 & 2697 & $27.11\pm0.10$ & 2797 & $26.07\pm0.07$ & ... & ...\\
422.1 & 57939.5 & 2697 & ... & 2797 & ... & ... & ...\\
 \hline
\end{tabular}
\label{table:phot_nir}
\end{small}
\end{minipage}
\end{table*}

\begin{table}
\begin{small}
\caption{Log of our late-time \textit{Chandra}/ACIS-S observations of GRB 160625B and GRB 160509A.}
\begin{tabular}{llccc}
\hline
Phase & MJD & $t_{\mathrm{exp}}$ & $f_{\nu}$(5 keV) \\
(d) & & (ks)  & (erg s$^{-1}$ cm$^{-2}$ keV$^{-1}$) \\
\hline
160625B \\
\hline
69.8 & 57634.7 & 19.80 & ($1.47 \pm 0.29) \times 10^{-15} $ \\
142.3 & 57707.3 & 45.84 & ... \\
$144.3 \pm 2.2$\footnote{Combination of the 142.3 and 146.2 d epochs.} & $57709.3 \pm 2.2$ & 69.56 & ($3.21 \pm 0.79) \times 10^{-16} $ \\
146.2 & 57711.2 & 23.72 & ... \\
\hline
160509A \\
\hline
42.1 & 57559.5 & 24.75 & ($1.38 \pm 0.25) \times 10^{-15} $ \\
 \hline
\end{tabular}
\label{table:xray}
\end{small}
\end{table}

GRB 160625B was observed in the radio using the VLA in the $C$, $K$, $X$ and/or $Ku$ bands at five epochs between 2016 March 30 (4.5 d) and 2017 January 20 (209.0 d), and GRB 160509A in the $C$ and $X$ bands on 2016 June 2 (23.9 d), 2016 June 15 (36.9 d) and 2016 July 28 (79.9 d) (program IDs S81171 and SH0753, PI Cenko and Fruchter respectively). The observations were done in the B configuration, apart from the last GRB 160625B point where configuration A was used. The log of our observations is presented in Table \ref{table:radio}. The data were reduced using the Common Astronomy Software Applications package \citep[CASA;][]{mcmullin07}\footnote{https://casa.nrao.edu}. Calibration was carried out using the standard VLA calibration pipeline provided in CASA. For GRB 160625B we used J2049+1003 as our complex gain calibrator and 3C48 as our flux and bandpass calibrator. For GRB 160509A we used J2005+7752 as our complex gain calibrator and 3C48 as our flux and bandpass calibrator. After calibration, the data were manually inspected for radio-frequency interference flagging. Imaging was carried out using the \texttt{clean} algorithm in interactive mode in CASA. Flux densities reported in Table \ref{table:radio} correspond to peak flux densities measured in a circular region centered on the GRB position, with radius comparable to the nominal full width half maximum of the VLA synthesized beam in the appropriate configuration and frequency band. The reported errors include the VLA calibration uncertainty, which is assumed to be 5 per cent below 18 GHz and 10 per cent above it\footnote{(\url{https://science.nrao.edu/facilities/vla/docs/manuals/oss/performance/fdscale})}.

\begin{table}
\centering
\begin{small}
\caption{Log of our VLA radio observations of GRB 160625B and GRB 160509A. The GRB 160625B points until 31.3 d were also reported in \citet{troja17}, but without the calibration uncertainty.}
\begin{tabular}{llccc}
\hline
Phase & MJD & $\nu$ & $f_{\nu}$ & Configuration \\
(d) & & (GHz) & ($\mu$Jy) & \\
\hline
160625B \\
\hline
4.5 & 57569.4 & 4.8 & $104 \pm 16$ & B \\
4.5 & 57569.4 & 7.4 & $454 \pm 27$ & B \\ 
4.5 & 57569.4 & 19 & $278 \pm 35$ & B \\ 
4.5 & 57569.4 & 25 & $204 \pm 36$ & B \\ 
13.4 & 57578.3 & 4.8 & $377 \pm 25$ & B \\
13.4 & 57578.3 & 7.4 & $310 \pm 21$ & B \\
13.4 & 57578.3 & 22 & $163 \pm 20$ & B \\
31.3 & 57596.2 & 7.4 & $113 \pm 16$ & B \\
31.3 & 57596.2 & 22 & $88 \pm 19$ & B \\
58.3 & 57623.2 & 6.1 & $75 \pm 11$ & B \\
58.3 & 57623.2 & 22 & $52 \pm 13$ & B \\
209.0 & 57773.9 & 6.1 & $16 \pm 5$ & A \\
\hline
160509A \\
\hline
23.9 & 57541.3 & 6.0 & $80 \pm 8$ & B \\
23.9 & 57541.3 & 9.0 & $71 \pm 7$ & B \\
36.9 & 57554.3 & 5.0 & $50 \pm 7$ & B \\
36.9 & 57554.3 & 6.9 & $52 \pm 7$ & B \\
36.9 & 57554.3 & 8.5 & $41 \pm 6$ & B \\
36.9 & 57554.3 & 9.5 & $29 \pm 6$ & B \\
79.9 & 57597.3 & 6.0 & $27 \pm 6$ & B \\
79.9 & 57597.3 & 9.0 & $25 \pm 5$ & B \\
 \hline
\end{tabular}
\label{table:radio}
\end{small}
\end{table}

\section{Analysis} \label{sec:analysis}

\subsection{GRB 160625B}

As our \emph{HST} observations took place after the jet break, we combined our data set with earlier ground-based observations. Both \citet{alexander17} and \citet{troja17} have published SDSS $r'$ band light curves of GRB 160625B. However, there is a slight ($\sim 0.1$ mag) systematic offset between these data, so in our light curve fits we have only used the \citet{troja17} data set, which has a larger number of data points and which was directly tied to the PanSTARRS magnitude system. Magnitudes of GRB~160625B in the $r'$ band were converted to flux density at the central wavelength of the F606W filter (5947~\AA) assuming a spectral slope of $f_{\nu} \propto \nu^{-0.68}$ between the characteristic synchrotron frequency $\nu_m$ and the cooling frequency $\nu_c$ \citep{alexander17}. As the optical spectrum with $\beta = -0.68\pm0.07$ is consistent with the expected index of $\beta = -0.65$ when $p = 2.3$ (also consistent with the light curve; see Section \ref{sec:disco0625b}), host extinction is assumed to be negligible. Optical fluxes have been corrected for Galactic reddening, $E(B-V) = 0.1107$ mag \citep{dustmap}, assuming the \citet{cardelli89} extinction law. In the X-ray, we combined our \textit{Chandra} data with the GRB 160625B light curve from the \textit{Swift}/XRT lightcurve repository\footnote{\href{http://www.swift.ac.uk/xrt_curves/}{http://www.swift.ac.uk/xrt\_curves/}} \citep{evans07,evans09}, converted to 5 keV flux densities using {\sc pimms} as described in Section \ref{sec:data}.

\begin{figure}
\centering
\includegraphics[width=1.0\columnwidth]{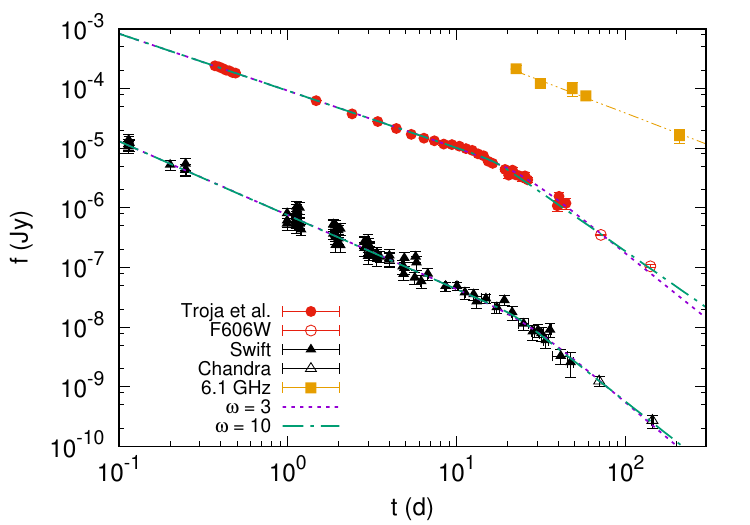} 
\caption{Observed optical (extinction-corrected), X-ray and interpolated 6.1 GHz light curve of the afterglow of GRB 160625B (points) and our power law fits including the broken power laws described by Eq. 1 (lines). The $r'$-band magnitudes from \citet{troja17} (solid circles) have been converted into flux density. X-ray flux densities from \textit{Swift}/XRT (solid triangles) and \textit{Chandra}/ACIS-S (open triangles) are reported at 5 keV. The post-break fit is better assuming $\omega = 10$ (dot-dashed green line), especially regarding the optical \textit{HST} point at 140.2 d. The pre-break fit does not depend on the choice of $\omega$.}
\label{fig:lc}
\end{figure}

We then fitted a smooth broken power law of the form
\begin{equation}
f_{\nu} = f_{\nu,0}\Big[\Big(\frac{t}{t_j}\Big)^{-\omega \alpha_{1}} + \Big(\frac{t}{t_j}\Big)^{-\omega \alpha_{2}}\Big]^{-\frac{1}{\omega}}
\end{equation}
to the light curve, where $t_j$ is the jet break time, $\alpha_{1}$ is the pre-break power-law slope, $\alpha_{2}$ the post-break slope, and $\omega$ describes the sharpness of the break. We fitted this function to both the optical and the X-ray curve using two values, 3 and 10, for $\omega$ \citep[a value of 3 was found consistent with most GRB observations by][but some events were found to require a sharper break with $\omega = 10$]{liang07}. The results of the fit parameters are presented in Table \ref{table:fits}. The pre-break decline $\alpha_{1}$ does not depend on the choice of $\omega$; we found $\alpha_{1,\mathrm{F606W}} = -0.96 \pm 0.01$ and $\alpha_{1,\mathrm{X}} = -1.24 \pm 0.02$ in both cases. The best fit to the post-break decline was $\alpha_{2,\mathrm{F606W}} = -2.27 \pm 0.13$ and $\alpha_{2,\mathrm{X}} = -2.40 \pm 0.19$ assuming $\omega = 3$, and $\alpha_{2,\mathrm{F606W}} = -1.96 \pm 0.07$ and $\alpha_{2,\mathrm{X}} = -2.23 \pm 0.15$ when $\omega = 10$. The optical and X-ray light curves and our best fits in both cases are shown in Figure \ref{fig:lc}. 

\begin{table}
\centering
\begin{small}
\caption{Parameters of the best smooth broken power law fits to the GRB 160625B light curves and of the single power law (SPL) fits to the early and late decline, ignoring the bump(s) between 8.5 and 26.5 d.}
\begin{tabular}{lccc}
\hline
Parameter & $\omega = 3$ & $\omega = 10$ & SPL\\
 \hline
 $t_{j,\mathrm{F606W}}$ & $24 \pm 3$ d & $17 \pm 2$ d & $17 \pm 4$ d\\
 $\alpha_{1,\mathrm{F606W}}$ & $-0.96 \pm 0.01$ & $-0.96 \pm 0.01$ & $-0.97 \pm 0.01$ \\
 $\alpha_{2,\mathrm{F606W}}$ & $-2.27 \pm 0.13 $ & $-1.96 \pm 0.07 $ & $-1.94 \pm 0.13$\\
 Reduced $\chi^2$ & 5.5 & 4.4 & 1.8\\
 \hline
 $t_{j,\mathrm{X}}$ & $27 \pm 5 $ d & $22 \pm 4 $ d & $22 \pm 5$ d\\
 $\alpha_{1,\mathrm{X}}$ & $-1.24 \pm 0.02$ & $-1.24 \pm 0.02$ & $-1.25 \pm 0.03$\\
 $\alpha_{2,\mathrm{X}}$ & $-2.40 \pm 0.19$ & $-2.23 \pm 0.15$ & $-2.20 \pm 0.13$ \\
 Reduced $\chi^2$ & 0.91 & 0.81 & 0.84\\
 \hline
\end{tabular}
\label{table:fits}
\end{small}
\end{table}

We also fitted the decline using a single power law before 8.5 d and another after 26.5 d, ignoring the points in the vicinity of the break itself. The $r'$ band light curve contains at least one smooth 'bump' feature, possibly two depending on $t_j$ (we discuss the nature of the bump in Section \ref{sec:shape}). These may disturb the optical broken power-law fits; the reduced $\chi^2$ values of these fits are rather high, although the small errors also contribute to this. The result is $\alpha_{2,\mathrm{F606W}} = -1.94 \pm 0.13$, nearly exactly coinciding with the $\omega = 10$ case but with a $\sim2.5 \sigma$ difference to $\omega = 3$. Repeating this in the X-ray results in $\alpha_{2,\mathrm{X}} = -2.20 \pm 0.13$, which is also almost identical to the $\omega = 10$ case. A simultaneous single power law fit to both post-break light curves results in $\alpha_{2} = -2.01 \pm 0.09$. 

Assuming an achromatic break, we determined $t_j$ by taking the weighted average of $t_{j,\mathrm{F606W}}$ and $t_{j,\mathrm{X}}$. In the $\omega = 10$ case, the result is $t_j = 19 \pm 2$ d. Assuming an instantaneous break (corresponding to $\omega = \infty$) between the single power law fits, the resulting jet break times are consistent, $t_{j,\mathrm{F606W}} = 17 \pm 4$ d and $t_{j,\mathrm{X}} = 22 \pm 5$ d, and the weighted average $t_j = 19 \pm 3$ d. In the $\omega = 3$ case, we obtained $t_j = 25 \pm 3$ d.

For the radio light-curve of GRB 160625B, we combined flux measurements from \cite{alexander17} and \citet{troja17} with our own data. At 58.3 d and 209.0 d we have observations at 6.1 GHz; we therefore obtained flux densities at 6.1 GHz by power-law interpolation between 5 and 7.1 GHz literature values at 22.5 and 48.4 d. We also scaled the 7.4 GHz flux at 31.34 d assuming the same power law as at 22.5 d. Points earlier than 22.5 d were ignored in the analysis of the late afterglow due to the influence of the reverse shock \citep{alexander17}. The resulting best fit for the late-time light curve is $\alpha_{\mathrm{6.1GHz}} = -1.08 \pm 0.11$ as shown in Figure \ref{fig:lc}. 

\begin{figure}
\includegraphics[width=\columnwidth]{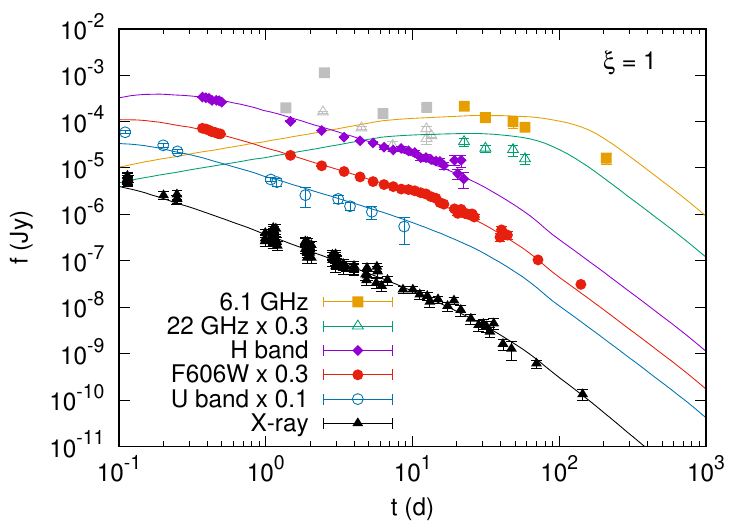} \\ \includegraphics[width=\columnwidth]{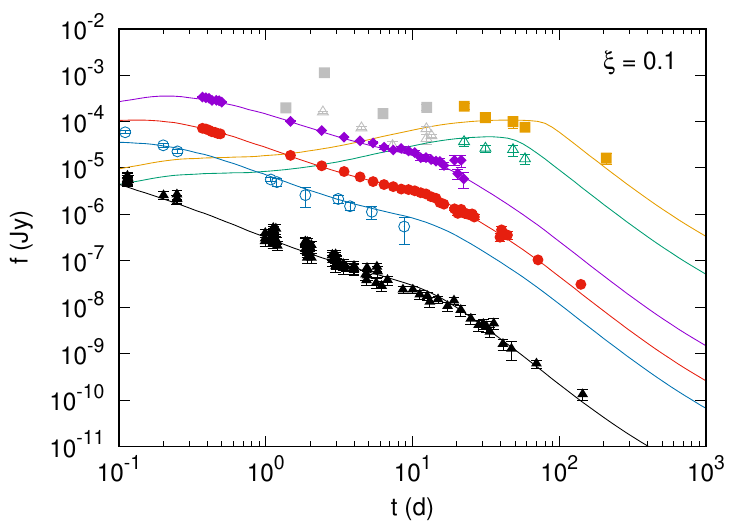} \\
\includegraphics[width=\columnwidth]{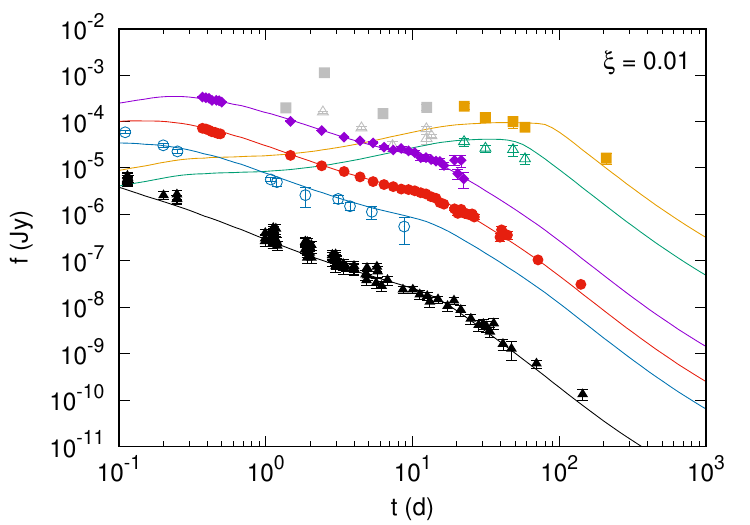} 
\caption{Observed X-ray, optical ($U$, F606W and $H$ bands shown here) and interpolated 6.1 and 22 GHz light curves of the afterglow of GRB 160625B (points), and the best fits given by {\sc boxfit} (lines) at indicated participation fraction $\xi$. The shape of the radio light curve is not well reproduced by any of the fits. Data denoted by grey points are ignored in the fitting (see text).}
\label{fig:lc_boxfit}
\end{figure}

Additionally, we used the {\sc boxfit} v.2 afterglow fitting code \citep{boxfit}, based on the Afterglow Library\footnote{http://cosmo.nyu.edu/afterglowlibrary/index.html}, to fit the light curve. The library of models itself was constructed using the relativistic hydrodynamics code {\sc ram} \citep{ram06}. {\sc boxfit} then uses a downhill simplex method with simulated annealing to find the best fit, interpolating between these models. We omitted the pre-break radio points due to the influence of the reverse shock in the early light curve, and all the radio points below 5 GHz due to possible strong Milky Way scintillation \citep{alexander17}. We also included the ultraviolet to near-infrared frequency data from \citet{troja17}. We assumed an ISM-like CBM (the light curve rules out a wind-type CBM; see Section \ref{sec:disco0625b}) and performed the fit with three different values of the participation fraction $\xi$, i.e. the fraction of electrons accelerated by the shock into a non-thermal power-law distribution. Simulations indicate this value can be as low as 0.01 \citep{sironi11,sironi13,warren18}; we used fixed values of 1 (commonly assumed in the literature), 0.1 and 0.01. All other model parameters were allowed to vary within the full range allowed by {\sc boxfit}. The resulting best-fit parameters are summarized in Table \ref{table:boxfit1}. Taking the isotropic-equivalent $\gamma$-ray energy $E_{\mathrm{iso}} = 3.0 \times 10^{54}$ erg \citep[with the fluence from][]{svinkin16}, we also calculate the geometry-corrected total energy and the efficiency $\eta = E_{\mathrm{iso}}/(E_{\mathrm{K,iso}} + E_{\mathrm{iso}})$ for the conversion of kinetic energy to $\gamma$-rays. These fits, however, fail to reproduce the measured power law slope of $\alpha_{\mathrm{6.1GHz}} = -1.08 \pm 0.11$, instead predicting a break in the radio light curve around $\sim100$~d (associated with the passage of $\nu_m$ through this band). See Figure \ref{fig:lc_boxfit} for our best {\sc boxfit} light curve fits. For clarity, we plot the $U$, F606W and $H$ bands, covering the optical/infrared behavior from early to late times, but omit the other optical/infrared bands, which exhibit very similar behavior \citep[see][]{troja17}. While the late-time 6.1 GHz light curve can be reproduced slightly better at low $\xi$ values, the fit at higher frequencies or earlier times is still somewhat worse; we show 22~GHz as an example.

\begin{figure}
\includegraphics[width=1.0\columnwidth]{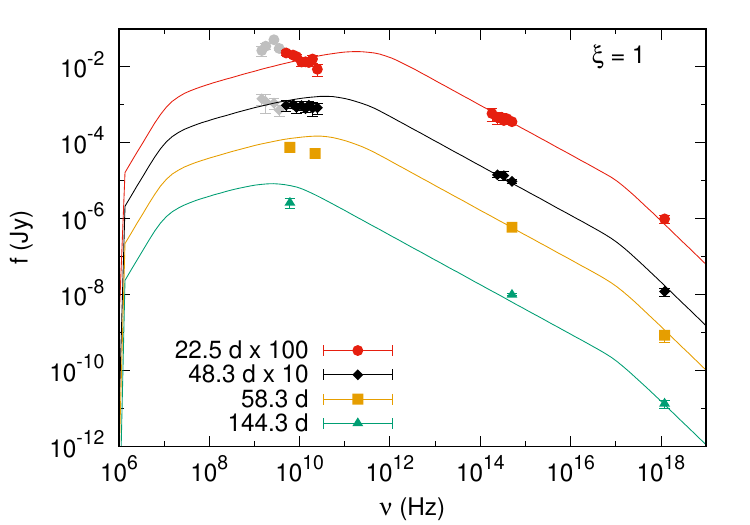} 
\includegraphics[width=1.0\columnwidth]{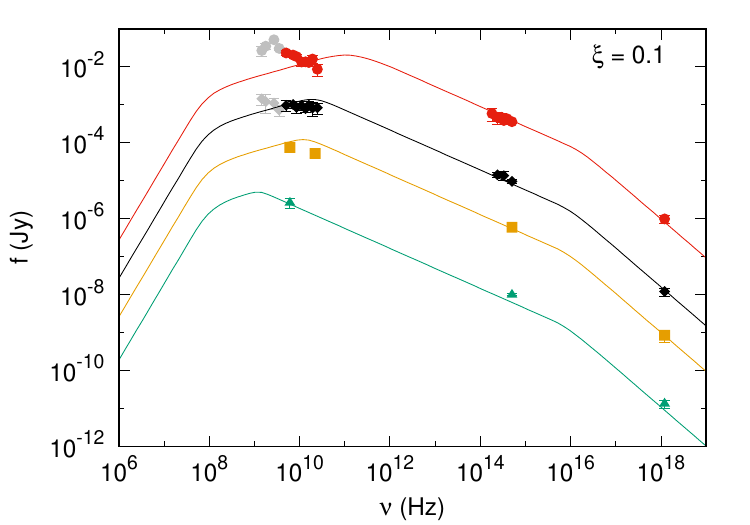}
\includegraphics[width=1.0\columnwidth]{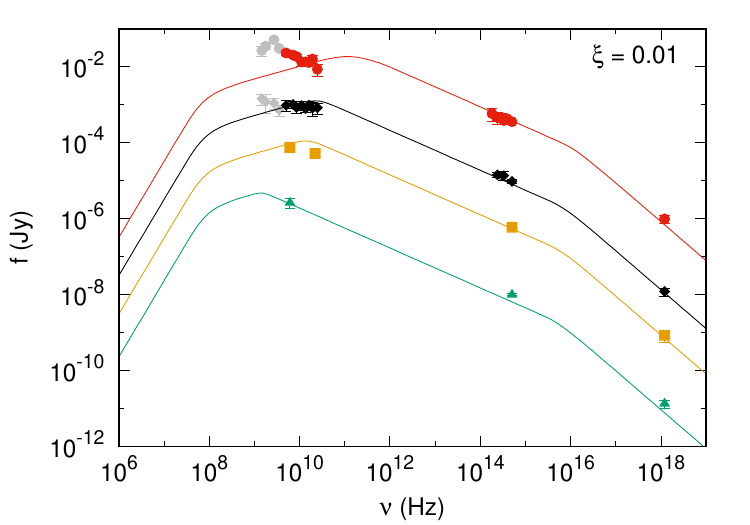} 
\caption{Observed spectral energy distribution of GRB 160625B (points) at late times, interpolated as necessary to the indicated dates, and the best fits given by {\sc boxfit} (lines) at indicated participation fraction $\xi$, using an constant CBM density profile. Data denoted by grey points are ignored in the fitting (see text).}
\label{fig:spec_boxfit}
\end{figure}

\begin{table}
\centering
\begin{small}
\caption{Best-fit physical parameters of the best {\sc boxfit} fits to GRB 160625B at three different values of the participation fraction $\xi$.}
\begin{tabular}{lccc}
\hline
Parameter & $\xi = 1$ & $\xi = 0.1$ & $\xi = 0.01$ \\ 
\hline
$p$ & 2.30 & 2.05 & 2.05 \\
$E_{\mathrm{K,iso}}$ (erg) & $1.8 \times 10^{54}$ & $1.4 \times 10^{54}$ & $1.3 \times 10^{55}$ \\
$\epsilon_e$ & 0.13 & 0.25 & 0.024 \\
$\epsilon_B$ & $0.030$ & $3.0 \times 10^{-4}$ & $5.8\times 10^{-5}$ \\
$n$ (cm$^{-3}$) & $1.1 \times 10^{-5}$ & 0.18 & 0.96 \\
$\theta_{j}$ (rad) & 0.059 & 0.14 & 0.13 \\
$\theta_{j}$ (deg) & $3.4$ & 7.8 & 7.2 \\
$\theta_{obs}$ (rad) & $0.012$ & $1.1\times10^{-3}$ & $1.1\times10^{-3}$ \\
$\theta_{obs}$ (deg) & $0.69$ & 0.07 & 0.06 \\
 \hline
$E_{\mathrm{tot}}$ (erg) & $8.3 \times 10^{51}$ & $4.1 \times 10^{52}$ & $1.3 \times 10^{53}$ \\ 
$\eta$ & 0.62 & 0.68 & 0.19 \\
$\chi^2 / \mathrm{d.o.f}$ & 8.6 & 4.6 & 4.5 \\
 \hline
\end{tabular}
\label{table:boxfit1}
\end{small}
\end{table}

\begin{figure}
\centering
\includegraphics[width=1.0\columnwidth]{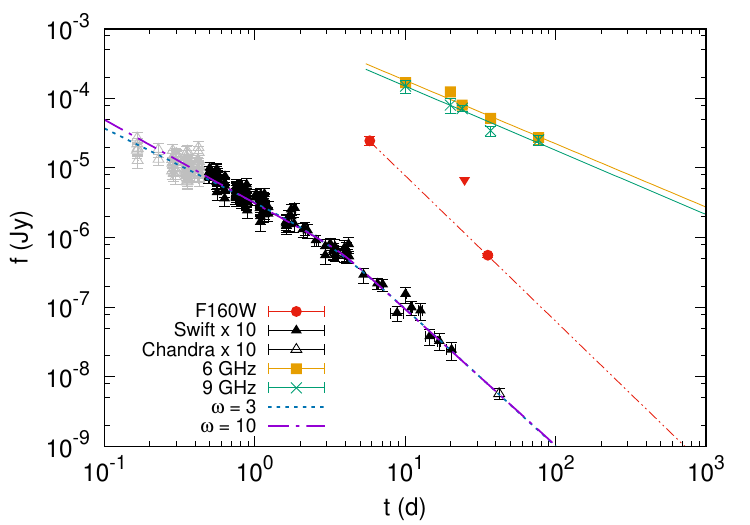} 
\caption{Observed F160W (extinction-corrected), X-ray and interpolated 6 and 9 GHz light curves of the afterglow of GRB 160509A (points) and our power law fits including the broken power laws described by Eq. 1 (lines). The red triangle is the upper limit of the F160W flux at 24.8 d. X-ray flux densities from \textit{Swift}/XRT (solid triangles) and \textit{Chandra}/ACIS-S (open triangles) are reported at 5 keV. Both choices of $\omega$ fit the late light curve equally well. The early light curve exhibits a shallower decay and another break, and thus points before $4 \times 10^4$ s (grey) are ignored.}
\label{fig:lc2}
\end{figure}

As some optical and X-ray observations are nearly contemporaneous, we can construct the spectral energy distribution (SED) of GRB 160625B. Figure \ref{fig:spec_boxfit} shows the SED at four epochs around or after the break, along with spectra produced by {\sc boxfit} at these epochs. The power-law slope of the SED, $\beta$, between the optical ($r$) and X-ray (5 keV) bands, steepens slightly over time, from $-0.79 \pm 0.02$ between 3 and 10 d to $-0.86 \pm 0.04$ at 141 d. This is steeper than $-0.65$, expected from $p \approx 2.3$ implied by the early optical and X-ray light curves (see Section \ref{sec:disco0625b}) for $\nu < \nu_c$, but shallower than $-1.15$, which is expected for $\nu > \nu_c$. \citet{alexander17} obtain an early X-ray spectral slope similar to this, $\beta_X = -0.86^{+0.09}_{-0.10}$, and explain this as $\nu_c$ being located just below the X-ray band. However, according to the UKSSDC \emph{Swift} Burst Analyser\footnote{http://www.swift.ac.uk/burst\_analyser/00020667/} the X-ray photon index $\Gamma_X$ (and thus the spectral slope in X-ray) does not significantly evolve over the first 30 d but stays around $\sim1.8$, after which the spectrum seems to flatten to $\Gamma_X \sim 1.1$. This feature may not be real, though, as the Burst Analyzer light curve deviates much more from a clean power law when this is used in flux calculation -- thus we assume a constant $\Gamma_X$\footnote{The post-break X-ray slope would not change by changing $\Gamma_X$ at the latest \emph{Swift} points, as \emph{Chandra} points would be affected equally -- but $t_{j,\mathrm{X}}$ could be delayed.}. If $\nu_c$ was initially just below X-ray and changed as $\nu_c \propto t^{-1/2}$, one would expect the spectrum to instead steepen over time to its $\nu \gg \nu_c$ value. We discuss this evolution further in Section \ref{sec:disco0625b}.

\subsection{GRB 160509A}

It was noted in \citet{laskar16} that the host galaxy of GRB 160509A contributes substantially to the optical and infrared photometry, and that the event occurred behind a significant amount of extinction in the host galaxy. In order to estimate the host galaxy extinction along the line of sight to the GRB, we removed the foreground Galactic reddening of $E(B-V) = 0.2519$ mag \citep{dustmap} using the \citet{cardelli89} law, and assumed a $f_{\nu} \propto \nu^{\beta}$ SED, where $\beta = -0.6$ \citep[consistent with $\nu < \nu_{c}$ and $p \approx 2.2$, determined based on the X-ray spectrum and light curve by][]{laskar16}. For the host, we assume the \citet{pei92} extinction law for the Small Magellanic Cloud (SMC), as both \citet{kann06} and \citet{schady12} found the extinction curve in the SMC consistent with their samples. We fitted the observed optical-infrared SED simultaneously at two epochs, corrected using this extinction curve, to find the required extinction correction to match $\beta = -0.6$. The GRB flux in the $g'$ band at 1 d was estimated by subtracting the observed flux at 28 d \citep[$g' = 25.39 \pm 0.12$;][]{laskar16} from the flux at 1.0 d \citep[$g' = 25.03 \pm 0.15$;][]{cenkogrb16}. The host is assumed to dominate at 28 d due to the flatness of the light curve even after the X-ray break. In the $J$ band, we subtracted the flux of the host galaxy measured in the \HST F110W filter (using a 1 arcsec aperture) from the flux at 1.2 d \citep[$J \approx 19.7$;][]{tanvir16}. The $r'$ band was not included in the SED, as the late and early fluxes are consistent within $1 \sigma$ \citep{cenkogrb16,laskar16}. Our F110W and F160W observations at 35.3 d made up the other epoch to be fitted simultaneously. The resulting host extinction is $A_V = 2.8 \pm 0.1$ mag in the rest-frame \citep[this is somewhat lower than the result obtained by][using an afterglow model fit where the host flux was a free parameter]{laskar16}. Using the \citet{pei92} law, the extinction correction in F160W (approximately $i$-band in the rest frame) is thus 1.5 mag. In the Milky Way, the adopted $N_{\mathrm{H,int}} = 1.52 \times 10^{22}$ cm$^{-2}$ would correspond to $A_V \approx 6.9$ mag \citep{guver09}, suggesting a low $A_V/N_H$ ratio for Milky Way standards but higher than that of most GRB hosts. This ratio is consistent with the $A_V$ vs. $N_H/A_V$ relation in \citet{kruhler11}. As in the case of GRB 160625B, we combined our \textit{Chandra} data of GRB 160509A with the data from the \textit{Swift}/XRT light curve repository converted to 5 keV flux densities.

The CIRCE $H$-band fluxes were converted to the narrower F160W filter assuming $\beta = -0.6$. The F160W and X-ray data and our power-law fits are presented in Figure \ref{fig:lc2}, and the parameters of the fits are listed in Table \ref{table:fits2}. For our power law fits we ignore the data points before $\sim0.5$ d ($4 \times 10^{4}$ s), as the early X-ray light curve may contain a plateau and/or a flare; see Figure \ref{fig:lc2}. In this case the smooth- and sharp-break scenarios give similar results: the best fit for the post-break decline for $\omega = 3$ is $\alpha_{2,\mathrm{X}} = -1.98 \pm 0.10$ and for $\omega = 10$, $\alpha_{2,\mathrm{X}} = -1.96 \pm 0.09$. The jet-break times, $3.2 \pm 0.9$ d and $3.7 \pm 0.8$ d, respectively, are consistent with each other as well.

\begin{table}
\centering
\begin{small}
\caption{Parameters of the best smooth broken power law fits to the GRB 160509A X-ray light curve.}
\begin{tabular}{lcc}
\hline
Parameter & $\omega = 3$ & $\omega = 10$ \\
 \hline
 $t_{j,\mathrm{X}}$ & $3.2 \pm 0.9 $ d & $3.7 \pm 0.8 $ d\\
 $\alpha_{1,\mathrm{X}}$ & $-1.06 \pm 0.10$ & $-1.20 \pm 0.06$ \\
 $\alpha_{2,\mathrm{X}}$ & $-1.98 \pm 0.10$ & $-1.96 \pm 0.09$ \\
 Reduced $\chi^2$ & 0.84 & 0.85 \\
 \hline
\end{tabular}
\label{table:fits2}
\end{small}
\end{table}

In the radio, we obtained the fluxes at 6 and 9 GHz at the epochs earlier than 79.9 d by power-law interpolation between observed fluxes -- our measurements at 36.9 d and those published in \citet{laskar16} at earlier times. We then fitted a single power law to the points where the reverse shock should no longer dominate the radio flux \citep[i.e. $\geq10$ days;][]{laskar16}. The resulting decline slopes are $\alpha_{\mathrm{6GHz}} = -0.91 \pm 0.11$ and $\alpha_{\mathrm{9GHz}} = -0.92 \pm 0.13$. Since the reverse shock may still be contributing a non-negligible fraction of the flux at 10 d, we also performed the fit without this epoch. The results are consistent but less constraining: $\alpha_{\mathrm{6GHz}} = -1.07 \pm 0.18$ and $\alpha_{\mathrm{9GHz}} = -0.92 \pm 0.21$. The slopes at other frequencies between 5 and 16 GHz, fitted from 10 to 20 d, are all consistent with these, ranging from $-0.80 \pm 0.10$ (7.4 GHz) to $-1.02 \pm 0.04$ (8.5 GHz). In F160W and/or $H$, we only have two points and an upper limit; therefore we simply measure the decline assuming a single power law. As the first point at 5.8 d is after the jet break time we obtained from the X-ray fit, there should be no significant deviation from a single power law. The measured decline is $\alpha_{2,\mathrm{F160W}} = -2.09 \pm 0.10$, consistent within $1\sigma$ with the X-ray decline.

\begin{figure}
\includegraphics[width=1.0\columnwidth]{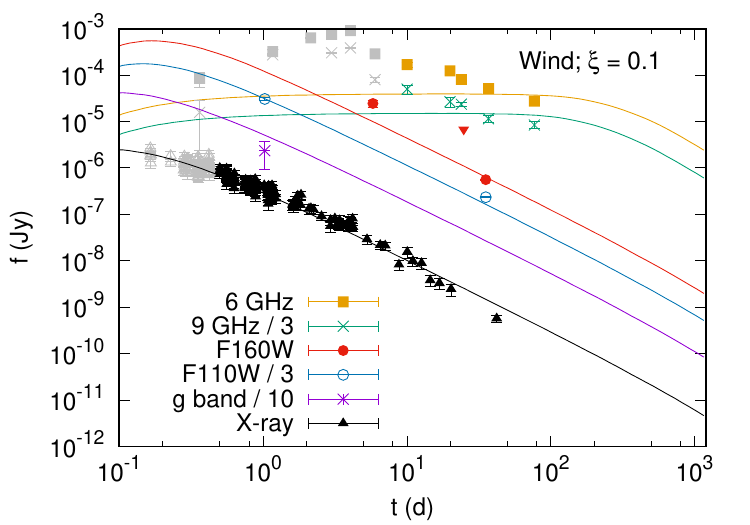} 
\caption{Observed X-ray and optical/infrared light curves and the interpolated 6 and 9 GHz light curves of the afterglow of GRB 160509A (points), and the best fit given by {\sc boxfit} (lines) using a wind-type CBM density profile and $\xi = 0.1$. The observed X-ray break is not reproduced (and indeed no break is seen even much later), and therefore a wind-type CBM is not considered further. Fits using $\xi=1$ and $\xi=0.01$ produce a similar light curve. Data denoted by grey points are ignored in the fitting (see text).}
\label{fig:windfit}
\end{figure}

\begin{figure}
\includegraphics[width=1.0\columnwidth]{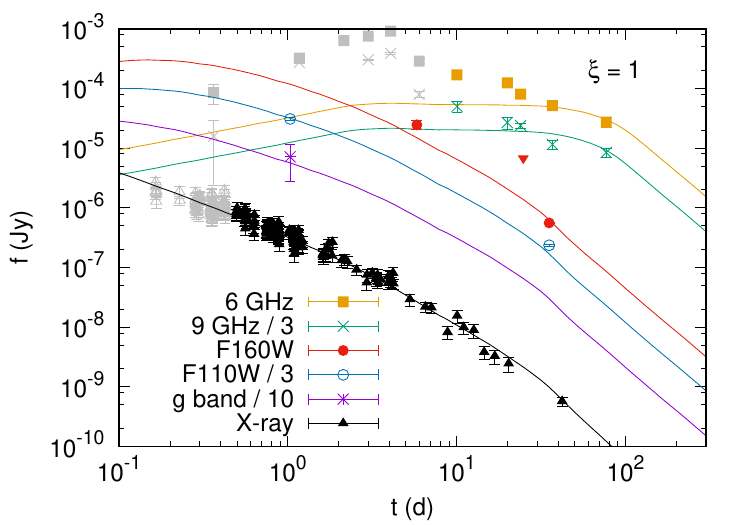} 
\includegraphics[width=1.0\columnwidth]{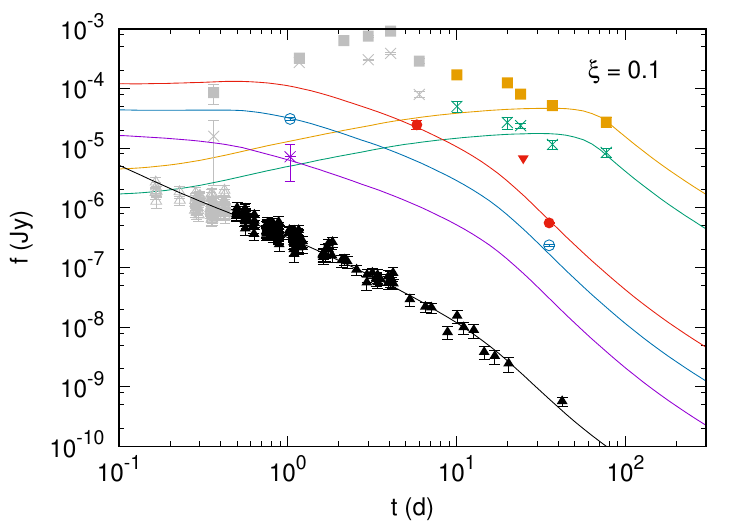}
\includegraphics[width=1.0\columnwidth]{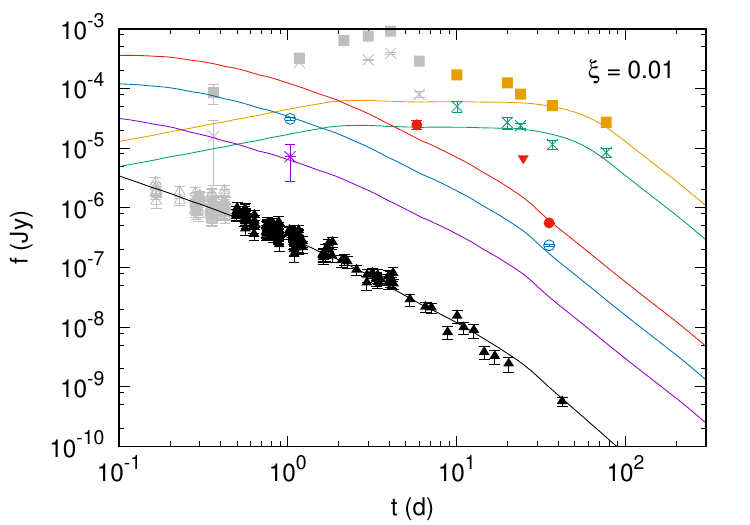} 
\caption{Observed X-ray and optical/infrared light curves and the interpolated 6 and 9 GHz light curves of the afterglow of GRB 160509A (points), and the best fits given by {\sc boxfit} (lines) at indicated participation fraction $\xi$, using an ISM-type CBM density profile. The radio light curve shape is again not well reproduced by the fits. Data denoted by grey points are ignored in the fitting (see text).}
\label{fig:lc2_boxfit}
\end{figure}

\begin{figure}
\includegraphics[width=1.0\columnwidth]{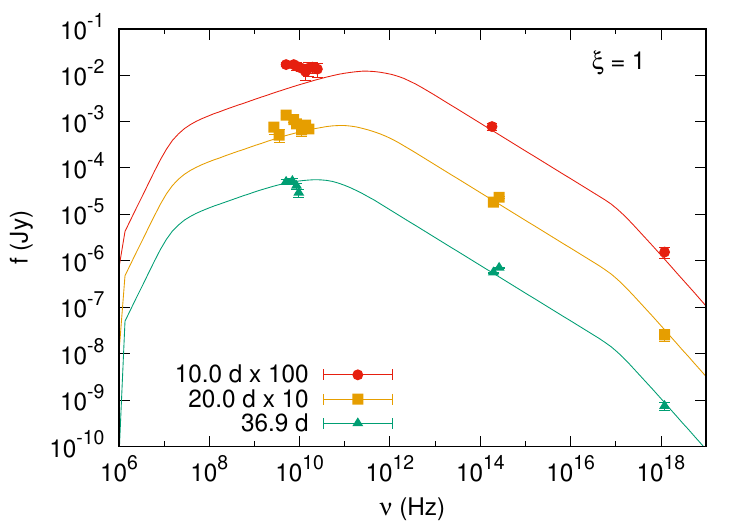} 
\includegraphics[width=1.0\columnwidth]{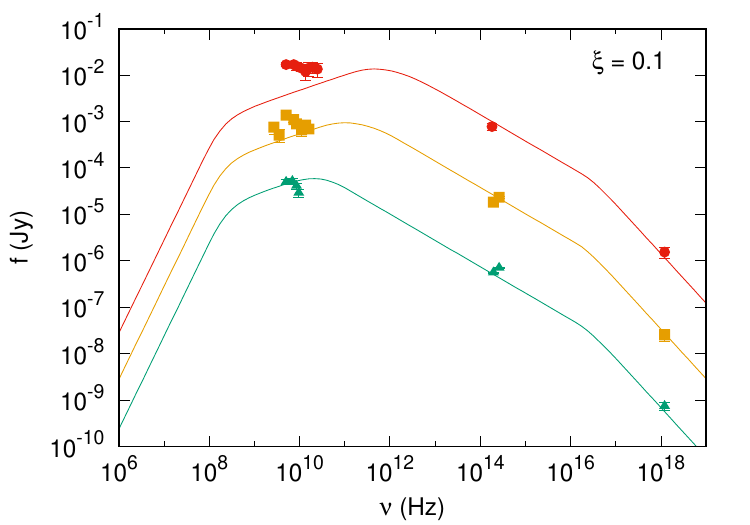}
\includegraphics[width=1.0\columnwidth]{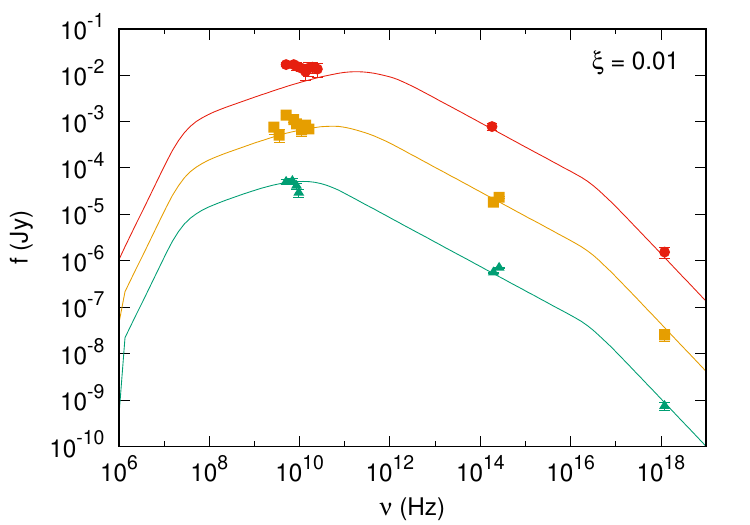} 
\caption{Observed spectral energy distribution of GRB 160509A (points) at late times, interpolated as necessary to the indicated dates, and the best fits given by {\sc boxfit} (lines) at indicated participation fraction $\xi$, using an ISM-type CBM density profile.}
\label{fig:spec2_boxfit}
\end{figure}

Using {\sc boxfit}, we again fitted the light curve at three different values of $\xi$: 1, 0.1 and 0.01. As with the power-law fits, the X-ray points before 0.6 d were ignored, since {\sc boxfit} cannot accommodate continuous energy injection. Radio points with a significant reverse shock contribution were also ignored \citep[i.e. $<10$ d; at frequencies $<5$ GHz also 10.03 d; see][]{laskar16}. We ran {\sc boxfit} with the boosted-frame wind-like CBM model (with both strong and medium boost) and a lab-frame model with ISM-like CBM, as the lack of optical data makes it difficult to distinguish between different CBM profiles \citep[although the ISM scenario is tentatively favored by][]{laskar16}. However, as shown in Figure \ref{fig:windfit}, our fits in a wind CBM do not reproduce the jet break clearly detected in the X-ray light curve. Even with the parameters in \citet{laskar16}, the break only appears at $\sim100$~d and the X-ray fit is much worse than with an ISM-type CBM. Thus the analytical model and {\sc boxfit} seem to disagree on how the jet behaves in a wind-type CBM, and we concentrate on the ISM scenario. The best ISM fits are shown in Figure \ref{fig:lc2_boxfit}; Figure \ref{fig:spec2_boxfit} shows the SED at three post-break epochs along with specra produced by {\sc boxfit} at these epochs. Our resulting best-fit parameters are summarized in Table \ref{table:boxfit2}. These fits (including the wind fits) again fail to match the observed shape of the radio light curve, although the amplitude of the flux can be reproduced at some epochs. 

\begin{table}
\centering
\begin{small}
\caption{Best-fit physical parameters of the best {\sc boxfit} fits to GRB 160509A at three different values of the participation fraction $\xi$.} 
\begin{tabular}{lccc}
\hline
Parameter & $\xi = 1$ & $\xi = 0.1$ & $\xi = 0.01$ \\ 
\hline
$p$ & 2.29 & $2.13$ & 2.05 \\
$E_{\mathrm{K,iso}}$ (erg) & $8.5 \times 10^{53}$ & $3.8 \times 10^{53}$ & $3.8 \times 10^{55}$ \\
$\epsilon_e$ & 0.19 & 0.45 & $5.7 \times 10^{-3}$\\
$\epsilon_B$ & 0.015 & $1.7 \times 10^{-5}$ & $5.8 \times 10^{-4}$\\
$n$ (cm$^{-3}$) & $2.1 \times 10^{-5}$ & $18.1$ & $6.1 \times 10^{-3}$\\
$\theta_{j}$ (rad) & 0.046 & 0.20 & 0.045 \\
$\theta_{j}$ (deg) & 2.6 & 11.5 & 2.6 \\
$\theta_{obs}$ (rad) & 0.026 & 0.12 & 0.027 \\
$\theta_{obs}$ (deg) & 1.5 & 7.0 & 1.5\\
 \hline
$E_{\mathrm{tot}}$ (erg) & $1.7 \times 10^{51}$ & $2.5 \times 10^{52}$ & $3.9 \times 10^{52}$ \\ 
$\eta$ & 0.50 & 0.69 & 0.02 \\
$\chi^2 / \mathrm{d.o.f}$ & 1.8 & 1.9 & 1.8 \\
 \hline
\end{tabular}
\label{table:boxfit2}
\end{small}
\end{table}

\section{Discussion} \label{sec:disco}

\subsection{The shape of the break}
\label{sec:shape}

In the X-ray, we find little difference in the reduced $\chi^2$ values of the fits between a sharp and a smooth break for GRB160625B. In the optical, however, fixing $\omega = 3$ results in a visible and significant residual of $4.2\sigma$ at 140.2 d, while fixing $\omega = 10$ results in a residual of $1.5\sigma$. The reduced $\chi^2$ of the latter fit is also slightly smaller. In the optical light curve, one can see either one slight bump or two, depending on the break time. These deviations from a perfect power law may disturb the fit and cause the high $\chi^2$ values, which suggests that one should also try only using the post-break points. Simply fitting a single power law to the points after 26.5 d results in consistency with the $\omega = 10$ case. We thus conclude that while both values of $\omega$ remain plausible, a sharp break with $\omega=10$ is more likely. A sharp break also implies a small viewing angle $\theta_{obs}$ \citep{ryan15}, which is compatible with the {\sc boxfit} results for this burst.

The post-jet-break decline of GRB 160625B has been previously estimated to be $f_{\nu} \propto t^{\alpha_2}$, where generally $\alpha_{2} \sim -2.3$ and its error roughly 0.5 \citep[][]{alexander17,fraija17, lu17}. These estimates are largely consistent with both sharp and smooth breaks (and with our results listed in Table \ref{table:fits}). However, all of these results are based on observations no later than $\sim 50$ d from the burst ($\sim2.5 \times t_j$, compared to our latest observations at $\sim7 \times t_j$), and their post-break fluxes mostly include relatively large uncertainties. In addition, \citet{troja17} obtained a more precise post-break slope of $\alpha_2 = -2.57 \pm 0.04$ and \citet{strausbaugh18} obtained $\alpha_{2,\mathrm{optical}} \approx 1.6$ and $\alpha_{2,\mathrm{X}} = -2.06 \pm 0.22$, but their optical slope is inconsistent with our later-time optical data in both cases. 

\citet{troja17} placed their estimate of the jet break at 14 d, during the 'bump' in the light curve between $\sim 8$ d and $\sim 16$ d. Using the same data, \citet{strausbaugh18} suggested a break at 12.6 d at the peak of the bump, which they took as brightening of the jet toward its edges. However, our later-time data require a later break and a steeper $\alpha_{2}$, leading us to suggest the bump may still be due to angular brightness differences or perhaps the result of density fluctuations in the CBM, but not necessarily a sign of a bright edge -- and seemingly not simultaneous with a true jet break. The bump is not seen in the X-rays, which is also consistent with a density fluctuation, as the flux above $\nu_c$ is insensitive to ambient density \citep{kumar00}. \citet{strausbaugh18} also suggest that a slowly changing spectral slope in the optical bands indicates a gradual cooling transition instead of a $\nu_c$ break in the spectrum, and that the optical spectrum eventually becomes consistent with $\beta \sim -1.1$, i.e. the slope above $\nu_c$, which would disfavor a CBM density fluctuation because of this insensitivity. We, however, measure $\beta = -0.86 \pm 0.04$ between F606W and 5 keV at 141 d, suggesting that $\nu_c$ is still above optical frequencies but below X-ray at this time. Thus we cannot rule out either scenario for the bump, but we can place the jet break at an epoch after the bump.

In the case of GRB 160509A, the $\chi^2$ values of the fits with different $\omega$ are close to equal and the post-break slopes are in agreement. A higher $\theta_{obs}$ results in a softer break \citep{ryan15}, so in this case, considering that $\theta_{obs} \sim 0.6 \theta_j$ (from {\sc boxfit}), one would expect the break to be softer than for GRB 160625B where $\theta_{obs}$ is much smaller or close to zero. One can attempt to resolve this by finding inconsistencies in estimates of $p$ based on the pre-break light curve and spectrum. The X-ray spectrum, with a slope of $\beta = -1.07 \pm 0.04$, is consistent with $p \approx 2.2$ and with $\nu_c$ being below the X-ray band \citep{laskar16}. As a result, we can use $\alpha = (2-3p)/4$ independent of the CBM distribution \citep{granotsari02}; in the case of $\omega = 3$ we obtain $p = 2.08 \pm 0.14$ and for $\omega = 10$, $p = 2.27 \pm 0.08$. While the former is closer to the measured post-break decline, both values are consistent with 2.2. 

\subsection{Physical implications}

\subsubsection{GRB 160625B} \label{sec:disco0625b}

Based on the well-constrained pre-break light curve of the afterglow of GRB 160625B, one can estimate the electron energy distribution index $p$: below the cooling frequency $\nu_c$, in the case of a wind-like CBM, $\alpha_{\mathrm{wind}} = (1-3p)/4$, while for a constant-density CBM similar to the interstellar medium (ISM), $\alpha_{\mathrm{ISM}} = 3(1-p)/4$ \citep{granotsari02}. Thus, in the optical, one obtains $p = 1.63 \pm 0.02$ in the wind case and $p = 2.29 \pm 0.02$ in the ISM case. Above $\nu_c$, in both cases $\alpha = (2-3p)/4$. Comparing the optical and X-ray spectra and fluxes \citet{alexander17} argue that $\nu_c$ lies below the X-ray frequencies after $\sim 1.2 \times 10^4$ s, and thus the early X-ray light curve gives us $p = 2.29 \pm 0.06$. This is also consistent with the spectrum below the X-ray frequencies \citep{alexander17}, and thus, as the $p$ values in the wind scenario are mutually inconsistent, an ISM-like density profile is favored. \citet{fraija17} infer a transition from wind-like to ISM-like CBM at $\sim 8000$ s.

When only taking into account the relativistic visible-edge effect \citep[][]{mrees99}, the slope of the decline is expected to steepen in the jet break by a factor of $t^{-3/4}$ in a constant-density CBM. In the $\omega = 10$ case, the difference between the pre- and post-break power laws is $\Delta\alpha_{\mathrm{F606W}} = -1.00 \pm 0.08$ in the optical and $\Delta\alpha_{\mathrm{X}} = -0.99 \pm 0.16$ in the X-ray. Thus a $t^{-3/4}$ factor can be ruled out in the optical at a $>3\sigma$ level (although in the X-ray, only at a $\sim 1.5\sigma$ level). The difference is larger in the $\omega = 3$ case ($>4 \sigma$ and $>2 \sigma$ respectively), and therefore a simple edge effect is inconsistent with our observations regardless of whether the break is sharp or smooth (the $t^{-1/2}$ factor from a wind-like CBM is, of course, even less plausible). 

If one assumes a smooth break ($\omega = 3$), both the optical and X-ray post-break decline rates are consistent with the form $f_{\nu} \propto t^{-p}$, for $p\approx2.3$, as expected from exponential lateral expansion \citep{rhoads99,sari99}. At first glance, the favored sharp-break scenario seems to make GRB 160625B inconsistent with a $f_{\nu} \propto t^{-p}$ decline in the optical band (the X-ray slope is still consistent with it) and would seem to require another physical mechanism. One explanation could be that the true jet break is due to a combination of the visible-edge effect and more limited lateral expansion. The steepening in both bands is a factor of $t^{-1}$, steeper than the $t^{-3/4}$ expected from the edge effect \citep[][]{mrees99,panmesz99}, and the resulting $\alpha_2$ values are only consistent within $2\sigma$, while the full exponential lateral expansion scenario described by \citet{rhoads99} should result in identical slopes. In some numerical simulations, lateral expansion has been found to initially involve only the outer layer of the jet carrying a fraction of its energy, and the bulk of the material remains unaffected for some time \citep{veerten12}, while the results of \citet{rhoads99} require the assumption that the entire jet expands at the speed of sound. On the other hand, completely ignoring the lateral expansion was found to result in insufficient steepening across the jet break. This scenario seems consistent with our results.

A complication was noted by \citet{gompertz18}, who find that using different synchrotron relations to estimate $p$ (such as using the spectral index or the pre- or post-break decline) typically results in different estimates, with an intrinsic scatter on the value of $p$ of $0.25 \pm 0.04$ (we will denote this as $\sigma_p$). They argue this is probably caused by emission from GRB afterglows not behaving exactly as the rather simplified analytical models predict\footnote{We note that the inconsistency between $p$ values derived from the optical and X-ray pre-break slopes assuming a wind-type CBM  is $>2 \sigma_p$, so an ISM-like density profile is still favored.}. Taking this scatter into account, both $\alpha_{2,\mathrm{F606W}}$ and $\alpha_{2,\mathrm{X}}$ in the $\omega = 10$ case (or simply using only the $>26.5$ d points and a single power law) are in fact consistent within $\approx1\sigma_{p}$ with $f_{\nu} \propto t^{-p}$. Thus lateral expansion at the speed of sound can still account for the observed late-time decline. Using closure relations for both the light curve and the spectrum, \citet{gompertz18} found a best fit of $p = 2.06 \pm 0.13$ for GRB 160625B, which is consistent with our results in both bands within $\sigma_p$. In any case, for this burst some form of lateral expansion is required, and the edge effect alone is insufficient. 

We can also attempt to use the results from {\sc boxfit} to determine if the magnetar spin-down power source is consistent with the GRB. The rotational energy that can be extracted from a millisecond magnetar is \citep{lu14,kumarzhang15}
\begin{equation}
    E_{\mathrm{rot}} \approx 2 \times 10^{52} ~\mathrm{erg} ~\frac{M}{1.4~M_{\odot}} \Big(\frac{R}{10~\mathrm{km}}\Big)^2 \Big(\frac{P_0}{1~\mathrm{ms}}\Big)^{-2}~,
\end{equation}
where $M$ is the mass, $R$ the radius and $P_0$ the initial spin period of the newborn magnetar. \citet{metzger15} placed a limit of $\sim1\times10^{53}$~erg on the maximum energy of a newborn magnetar in extreme circumstances (in terms of mass and spin period). Therefore the energy requirements of all the fits from {\sc boxfit} may technically be achievable with the magnetar model, but with the (more realistic) low $\xi$ values the required energy approaches or exceeds even this maximum limit. The exceptionally high $E_{\mathrm{iso}}$ can be due to a relatively narrow jet and a lower explosion energy instead, but this requires a high $\xi$ that is inconsistent with simulations by \citet{sironi11} and \citet{warren18} -- the best fit at $\xi =1$ also results in an extremely low density more typical to intergalactic environments. We do point out a caveat that the parameters of the best fits show a non-monotonic dependence on $\xi$, with notable degeneracy between parameters. 

We have attempted to use {\sc boxfit} to estimate errors for the best-fit parameters as well. However, as a result of what seems to be a bug in the error estimation routine of {\sc boxfit} (G. Ryan and H. van Eerten, private communication), some of the errors are clearly incorrect and, therefore, we have not included errors in our Table \ref{table:boxfit1}. This mostly manifests as error limits that either do not include the best fit or where the best-fit value of a parameter is always the lower limit as well\footnote{In other cases, such as the $\xi = 0.1$ case of GRB 160625B, the errors are \textit{seemingly} reasonable ($p = 2.05\pm0.01$, $E_{\mathrm{K,iso}} = 1.4_{-1.3}^{+1.2}\times10^{54}$~erg, $\epsilon_e = 0.25_{-0.13}^{+0.10}$, $\epsilon_B = 3.0_{-2.0}^{+106.3}\times10^{-4}$, $n = 0.18_{-0.15}^{+0.58}$~cm$^{-3}$, $\theta_j = 0.14\pm0.03$~rad and $\theta_{obs} = 1.1_{-1.1}^{+5.9}\times10^{-3}$~rad) and the relative ranges of each parameter comparable to those found by \citet{alexander17}. These values thus give an indication of how well each parameter is constrained.}. We also note that, as the shape of the radio light curve is not well reproduced in any of our fits, error limits could be misleading in any case. As a consistency check for the rest of the code, we have run {\sc boxfit} using the \citet{alexander17} forward shock parameters, which are similar to our $\xi=1$ results. The output light curves and spectra are similar to the analytical ones and reproduce the early behavior of the afterglow well, although post-break fluxes are somewhat under-predicted.

We also note that the 6.1 GHz light curve of GRB 160625B is not successfully reproduced by {\sc boxfit}, and the jet model struggles to explain the late slope of $\alpha_{\mathrm{6.1 GHz}} = -1.08 \pm 0.11$ and the lack of an observed jet break. At low $\xi$ values, the {\sc boxfit} fit is somewhat better, but only if one ignores the 22.5 d point, where a low-frequency scattering event by an intervening screen, suggested by \citet{alexander17}, may contribute to the flux. The radio SED shows a peak centered at 3 GHz between 12 and 22 d, which then disappears. Even so, the fit at 22 GHz is worse at all $\xi$ values. At 48 d, the radio SED is consistent with being entirely flat, which is only plausible in the standard model around a very smooth $\nu_m$ break. While the low $\xi$ fits do place the $\nu_m$ passage at roughly this time, the peak in the {\sc boxfit} spectrum is too sharp, and in earlier spectra the lowest frequencies must then be brightened by a factor of ten or so by the proposed scattering. The shape may instead be altered by another emission source contributing to the spectrum (see below).

Theoretically expected post-break values in the slow-cooling scenario ($\nu_m < \nu_c$) are $-p$ or $-1/3$, depending on which side of $\nu_m$ the band is located \citep{rhoads99}. As the jet break is a geometric effect, we should see it in every band, but this is not the case: we can set a limit of $t_{j,\mathrm{6.1GHz}} \gtrsim 10 \times t_{j,\mathrm{F606W}}$. The possibilities given by the standard jet model that are consistent with the slope are:
\begin{itemize}
    \item Post-break, $\nu_c < \nu_m$, i.e. fast-cooling: $\alpha_{\mathrm{6.1 GHz}}$ is consistent with the expected decline of $\alpha_2 = -1$. However, the measured $\alpha_{1,\mathrm{F606W}} = -0.96 \pm 0.01$ does not match the \emph{pre}-break decline expected at any frequency in this scenario.
    \item Pre-break, $\nu_m < 6.1~\mathrm{GHz} < \nu_c$: $\alpha_{\mathrm{6.1 GHz}}$ is consistent with $p = 2.4$ and $\alpha = 3(1-p)/4 = -1.05$ \citep{granotsari02}. However, the spectral index between radio and optical is $-0.35 \pm 0.03$ at 22 d and $-0.49 \pm 0.01$ at 140 d, which is intermediate between the indices expected above and below $\nu_m$ (respectively, $(1-p)/2 \approx -0.65$ and $1/3$) and thus implies that $\nu_m > 6.1$ GHz at 140~d, or that $p\approx2.0$.
    \item A transition to a non-relativistic flow, $\nu_m < 6.1~\mathrm{GHz} < \nu_c$: the expected slope is $(21-15p)/10$ \citep{vdhthesis}, resulting in $p=2.12\pm0.08$, which is consistent with our estimate within $\sigma_p$. However, such a transition is not seen in the optical or X-ray bands.
\end{itemize}

The LGRB population has been observed to be comprised of a radio-quiet and a radio-loud population, where the radio-quiet GRBs are incompatible with a simple sensitivity effect and indicate an actual deficit in radio flux compared to theory \citep{hancock13}. \citet{ronning18} further argued that the two populations originate in different progenitor scenarios. This deficit in radio flux implies some mechanism that suppresses the expected synchrotron emission at radio frequencies. Since our findings indicate that the radio light curve of GRB~160625B (and GRB 160509A; see below) is incompatible with the higher frequencies, the source of the radio emission that we do see may not be the same as that of the optical and X-ray synchrotron emission. This seems to suggest that even in (at least) some radio-loud GRBs, the same mechanism may be in effect. Furthermore, if the radio emission is generated by another source, this source is not active in the radio-quiet GRBs for some reason. We have run the {\sc boxfit} fitting code with $\xi=1$ and all radio fluxes divided by ten to investigate if the standard model allows suppression of the radio flux simply through adjusting the parameters. The resulting best fit over-predicts all radio fluxes by at least a factor of a few at all times. This implies a caveat that, at least in some cases, including another, dominant radio source without an additional suppression mechanism may over-predict the radio flux. Another caveat with this is that, unless the second component is coupled to the 'main' source, getting a total radio flux compatible with one component may require fine-tuning. If such a mechanism is widespread, one would expect some GRBs to have radio fluxes unambiguously too high for the standard model, which, to our knowledge, has not been seen.

One explanation for the 'extra' radio source, with its lack of a jet break and the requirement of $6.1~\mathrm{GHz} > \nu_m$, could be a two-component jet, where a narrow jet core is surrounded by a cocoon with a lower Lorentz factor \citep{berger03,peng05}, resulting in a different source with different physical parameters dominating the radio emission, and thus a different break time and $\nu_m$. This does not result in a deficit in radio synchrotron flux, only an inconsistency between the light curve shape and the standard model. For an on-axis or slightly off-axis burst ($\theta_{obs} < \theta_{j,\mathrm{narrow}}$), the wider component would not contribute significantly to the optical light curve if its kinetic energy is lower than that of the narrow component \citep{peng05}. This may also affect the required energy, but without robust modeling it is difficult to say whether the consistency with a magnetar energy source would change. 

\cite{strausbaugh18} suggested a scenario where a very smooth cooling transition (i.e. not a normal spectral break) is moving through the optical and infrared frequencies, starting at early times, and the optical spectrum becomes consistent with $\nu > \nu_c$ by $\sim50$ d. This would indicate a unique cooling behavior inconsistent with the standard expectations. The observed lack of evolution of the \emph{Swift} spectra until 30 d implies that the X-ray spectral slope $\beta_X$ is not the result of a $\nu_c$ break right below the X-ray frequencies, as this would require the spectrum to soften over time to its slope at $\nu \gg \nu_c$. Furthermore the optical-to-X-ray index is observed to gradually steepen and eventually become similar to $\beta_X$. This is qualitatively consistent with the reddening of the optical spectrum noted by \citet{strausbaugh18}. In addition, $\beta_X$ indicates a different $p$ than the X-ray light curve; this agrees with the implication of \citet{gompertz18} that some physics is missing or simplified in the relevant closure relations. Another possible explanation is that a Klein-Nishina correction \citep{nakar09} is needed above $\nu_c$; this can result in $\beta = 3(1-p)/4$, which would imply $p\approx2.1$. This harder spectrum is expected to dominate when the $\epsilon_e/\epsilon_B$ ratio is high, which would fit the low-$\xi$ {\sc boxfit} results.

\subsubsection{GRB 160509A} \label{sec:disco0509a}

In the case of GRB~160509A, the change in X-ray decay slope across the break, $\Delta\alpha_{\mathrm X} = -0.75 \pm 0.11$ for a sharp break and $\Delta\alpha_{\mathrm X} = -0.92 \pm 0.15$ for a smooth break. Thus we cannot exclude the $t^{-3/4}$ factor expected from the edge effect alone in an ISM-like medium. The $t^{-1/2}$ factor expected in the case of a wind medium is inconsistent with the observations at a $2.3\sigma$ or $3\sigma$ level, depending on $\omega$. However, when considering the intrinsic $p$ scatter of $\sigma_p = 0.25$ \citep{gompertz18}, $\alpha_{2,{\mathrm X}}$ is also consistent with a $f_{\nu} \propto t^{-p}$ decline. Thus we cannot say conclusively whether lateral expansion is important in the jet of GRB 160509A, but it does not seem \emph{necessary}. In the IR, the measured slope of $\alpha_{2,\mathrm{F160W}} = 2.09 \pm 0.10$ is marginally consistent ($1.1\sigma$) with $p \approx 2.2$, but a lack of pre-break data prevents us from determining $\Delta\alpha_\mathrm{F160W}$.

The decline of the afterglow in the radio after 10 d is about $f \propto t^{-0.9}$ at both 6 and 9 GHz (and consistent with this at other frequencies where fewer points are available). This is again inconsistent with the expected post-jet-break slope of $-p$ or $-1/3$ in the slow-cooling case, respectively above and below the characteristic synchrotron frequency $\nu_m$ \citep{rhoads99}. As with GRB 160625B, we list the possibilities consistent with this decline, allowed by standard jet theory:
\begin{itemize}
    \item Post-break, $\nu_c < \nu_m$, i.e. fast-cooling: $\alpha = -1$ is expected and consistent with $\alpha_{\mathrm{radio}}$, but this scenario is incompatible with the measured IR-to-X-ray spectral index $-0.74 \pm 0.09$ at 35 d, as the expected index is $-p/2 \approx -1.1$ (a photon index consistent with this is indeed seen in the X-ray at earlier times according to the UKSSDC \emph{Swift} Burst Analyser\footnote{http://www.swift.ac.uk/burst\_analyser/00020607/} -- $<\beta_X> = 1.06 \pm 0.04$ between 1 and 10 d -- indicating that $\nu_c$ is still between X-ray and optical frequencies and $\nu_c > \nu_m$ at 35~d).  
    \item Pre-break, $\nu_m < 6~\mathrm{GHz} < \nu_c$, ISM-like CBM: $\alpha_{\mathrm{radio}}$ is consistent with the expected decline \citep[$\alpha = 3(1-p)/4 = -0.9$ assuming $p = 2.2$;][]{granotsari02}, but the observed spectral index of $-0.40 \pm 0.01$ between F160W and 9 GHz at 35 d implies $\nu_m > 9$ GHz.
    \item A transition to a non-relativistic flow, $\nu_m < 6~\mathrm{GHz} < 9~\mathrm{GHz} < \nu_c$: the expected slope is $(21-15p)/10$, resulting in $p=2.01\pm0.08$ -- again consistent with our estimate within $\sigma_p$. However, such a transition is not seen in the X-ray light curve, which continues to evolve consistently with a relativistic flow.
\end{itemize}

The best {\sc boxfit} fit at $\xi=1$ places a smooth, and thus off-axis, jet-break at a later time, around 35 d in all bands, in which case the radio light curve would include contamination from the reverse shock at early times, changing the decline slope (see Figure \ref{fig:lc2_boxfit}). This is because {\sc boxfit} attempts to fit a model with a late break to $f_{\nu} \propto t^{-p}$ in order to match the radio light curve, which has no observed break. It is incompatible with the broken power-law fit with $t_j \sim 3.5$ d, though, and at lower, more realistic values of $\xi$ the break is placed at an earlier time. This scenario is therefore not supported. Instead, for GRB 160509A we can place a lower limit of $t_{j,\mathrm{radio}} \gtrsim 20 \times t_{j,\mathrm{X}}$ based on the broken power-law fit. The situation in the radio frequencies is thus qualitatively very similar to that of GRB 160625B, and the same mechanisms may well be in effect.  

We note that \citet{kangas19b} are, in fact, able to get a plausible fit to the GRB 160509A radio light curve using an analytical fit based on the standard model, but only if the light curve smoothly turns over to a $t^{-p}$ decline immediately after the last radio epoch, which is suspicious as their sample contains several GRBs with no unambiguously observed radio breaks, and many cases where the standard model does not fit the radio light curve.  We also note that as \citet{laskar16} showed, the radio SED seems to remain roughly flat after the reverse shock influence on the light curve fades ($\sim20$ d), which might again be caused by another emission component. As {\sc boxfit} also disagrees with this analytical model, one or the other is in doubt. The issue will be addressed in more detail in the upcoming revised version of that paper. 

A {\sc boxfit} simulation using the FS parameters of the \citet{laskar16} analytical model agrees fairly well with the X-ray data and reproduces the rough magnitude of the radio light curve but not its shape (assuming some RS contribution not accounted for by {\sc boxfit}), but over-predicts the IR flux by a factor of about 10. Their IR light curve does not include host subtraction, and they fit for extinction as another free parameter in their model. Our host subtraction allows us to estimate the extinction and true IR fluxes independently, and in light of this the \citet{laskar16} model becomes incompatible with the IR data. Thus our {\sc boxfit} results provide a better reproduction of the light curve in the IR. However, again, the fit parameters show a non-monotonic dependence on $\xi$. As with GRB 160625B above, {\sc boxfit} was clearly unable to produce meaningful error bars for the parameters in some cases, and these are not included in Table \ref{table:boxfit2}\footnote{Again, the ranges of each parameter at $\xi=0.1$, which are large but not obviously incorrect, may provide some indication of how well each parameter is constrained: ($p = 2.13_{-0.01}^{+0.02}$, $E_{\mathrm{K,iso}} = 3.8_{-3.4}^{+24.8}\times10^{53}$~erg, $\epsilon_e = 0.45_{-0.20}^{+0.31}$, $\epsilon_B = 1.7_{-0.7}^{+2.7}\times10^{-5}$, $n = 18_{-18}^{+1530}$~cm$^{-3}$, $\theta_j = 0.20_{-0.16}^{+0.18}$~rad and $\theta_{obs} = 0.12_{-0.12}^{+0.21}$~rad).} -- and, as the radio light curve is again problematic for the fit, would be misleading in any case.

Keeping in mind the caveats associated with our best {\sc boxfit} fits, we can use them to estimate the energy requirements. The geometry-corrected jet energy $1.8 \times 10^{51}$ erg at $\xi=1$ is well below the maximum rotational energy of a millisecond magnetar (see Section \ref{sec:disco0625b}) Once again, we deem the lower $\xi$ values more realistic based on simulations \citep{sironi11,warren18} and the fact that $\xi=1$ results in an extremely low density. Low $\xi$ values require energies around $\sim3\times10^{52}$ erg, which again strains the magnetar spin-down model but does not rule it out. Thus GRB 160509A also seems compatible with a magnetar power source.

For both GRBs considered here (Tables \ref{table:boxfit1} and \ref{table:boxfit2}), but especially for GRB 160509A, the efficiency $\eta$ of the prompt $\gamma$-ray emission depends on the value of $\xi$ used in the fitting, but not monotonically: with $\xi=0.01$ one obtains a much lower value for $\eta$ than otherwise. In both cases, the difference in $\chi^2$ between the $\xi=0.1$ and $\xi=0.01$ fits is minimal, and in the case of GRB 160509A, so is the difference between $\xi=1$ and $\xi=0.01$; thus, we cannot reliably distinguish between these scenarios. In the literature, it is commonly assumed that $\xi=1$, and high values of $\eta$ are obtained: for example, \citet{ronning04} find values as high as $\eta\sim1$ depending on $E_{iso}$, and mostly $\eta\gtrsim0.3$. Such a high efficiency is used as a criterion for successful models of prompt emission: e.g. the internal shock mechanism tends to result in $\eta\lesssim0.1$ \citep[][and references therein]{kumarzhang15}. Our results may indicate that, if very low values of $\xi$ are more realistic \citep{warren18}, one should not dismiss models based on low efficiency.

\section{Conclusions} \label{sec:concl}

We have presented our late-time optical, radio and X-ray observations of the afterglows of GRB 160625B and GRB 160509A. We have fitted broken power law functions to the data, combined with light curves from the literature, to constrain the jet break time and the post-jet-break decline, and used the numerical afterglow fitting software {\sc boxfit} \citep{boxfit} to constrain the physical parameters and energetics of the two bursts. Our conclusions are as follows:

Regardless of the sharpness of the GRB 160625B jet break, we find that the effect of the jet edges becoming visible as the jet decelerates is alone insufficient to explain the post-jet-break light curves. A full lateral expansion break onto a $t^{-p}$ decline is also inconsistent with the favored sharp break. The light curve behavior seems qualitatively consistent with the edge effect combined with only a fraction of the jet expanding at the speed of sound \citep{veerten12}. It is also possible that an intrinsic scatter in the electron Lorentz factor distribution index $p$ exists, the result of simplified synchrotron theory and closure relations that do not necessarily reflect the true complexity of the emission region \citep{gompertz18}. This scenario combined with lateral expansion is also consistent with our results. For GRB160509A we are unable to exclude any of the considered scenarios due to the scarcity of the available data.

Based on the best fits from {\sc boxfit}, the geometry-corrected energy requirements of both GRBs are consistent with a magnetar spin-down energy source -- albeit only in extreme cases when the 'participation fraction' (fraction of electrons accelerated into a non-thermal distribution) is fixed at $\xi=0.1$ or $\xi = 0.01$, requiring energies of $\sim3\times10^{52}$ or even $\sim10^{53}$ erg. As simulations have shown these lower fractions to be more realistic \citep[e.g.][]{warren18}, it seems that magnetar spin-down alone struggles to produce the required energies unless the nascent magnetar has extreme properties \citep{metzger15}. 

However, neither {\sc boxfit} nor analytical relations from standard jet theory \citep[e.g.][]{rhoads99,granotsari02} can provide a good fit to the radio data of either GRB, which are consistent with a single power law that requires the jet break to occur much later in radio than in the other bands. Both GRBs also show an almost flat radio SED at relatively late times \citep[tens of days; see][]{laskar16,alexander17}. The higher frequencies do conform to expectations from the jet model, though. This might be the result of a multi-component jet, but that would require the wide component of the jet to dominate the light curve, and simultaneously suppressed flux from the narrow component. A similar behavior (a radio decline described by a single power law with $\alpha = -1.19\pm0.06$ until $\sim60$ d) was recently reported for GRB 171010A by \citet{bright19}. We explore this problem further in a companion paper \citep{kangas19b}, and find that these GRBs are not exceptional in this regard.
\\

\acknowledgments

We thank the anonymous referee for comments that improved the paper. We also thank Hendrik van Eerten, Geoffrey Ryan, Alexander van der Horst, Paz Beniamini and Chryssa Kouveliotou for helpful discussions.

A.Co. acknowledges support from the National Science Foundation via CAREER award \#1455090. A.dU.P. acknowledges support from a Ramon y Cajal fellowship (RyC-2012-09975), from the Spanish research project AYA2017-89384-P, and from the State Agency for Research of the Spanish MCIU through the "Center of Excellence Severo Ochoa" award for the Instituto de Astrof\'isica de Andaluc\'ia (SEV-2017-0709). A.P. wishes to acknowledge support by the European Research Council via the ERC consolidating grant \#773062 (acronym O.M.J.). A.Cu. acknowledges the support of NASA MIRO grant NNX15AP95A.

Based on observations made with the NASA/ESA \textit{Hubble Space Telescope} (programme GO 14353, PI Fruchter), obtained through the data archive at the Space Telescope Science Institute (STScI).  STScI is operated by the Association of Universities for Research in Astronomy, Inc. under NASA contract NAS 5-26555. Support for this work was also provided by the National Aeronautics and Space Administration through Chandra Award Number 17500753, PI Fruchter, issued by the Chandra X-ray Center, which is operated by the Smithsonian Astrophysical Observatory for and on behalf of the National Aeronautics Space Administration under contract NAS8-03060. 

This work made use of data supplied by the UK Swift Science Data Centre at the University of Leicester as well as observations made with the Gran Telescopio Canarias (GTC), installed in the Spanish Observatorio del Roque de los Muchachos of the Instituto de Astrof\'{i}sica de Canarias, in the island of La Palma. Development of CIRCE was supported by the University of Florida and the National Science Foundation (grant AST-0352664), in collaboration with IUCAA.

The National Radio Astronomy Observatory is a facility of the National Science Foundation operated under cooperative agreement by Associated Universities, Inc.

Development of the {\sc boxfit} code was supported in part by NASA through grant NNX10AF62G issued through the Astrophysics Theory Program and by the NSF through grant AST-1009863. Simulations for {\sc boxfit} version 2 have been carried out in part on the computing facilities of the Computational Center for Particle and Astrophysics (C2PAP) of the research cooperation "Excellence Cluster Universe" in Garching, Germany.

%

\vspace{5mm}
\facilities{\textit{HST}(WFC3), \textit{Chandra}(ACIS-S), VLA, GTC(CIRCE)}


\software{{\sc drizzlepac} \citep{hack13}; {\sc pyraf} \citep{pyraf}; {\sc iraf} \citep{iraf}; {\sc ciao} \citep{ciao}; {\sc casa} \citep{mcmullin07}; {\sc isis} \citep{alard,alardlupton}; {\sc boxfit} \citep{boxfit}
          }

\bibliographystyle{aasjournal}
 \bibliography{references}





\end{document}